\documentstyle[aps,prb,preprint,tighten,epsfig]{revtex}

\def\bra#1{{\langle #1 |}}
\def\ket#1{{| #1 \rangle}}
\def\bracket#1#2{{\langle #1 | #2 \rangle}}
\def\expect#1{{\langle #1 \rangle}}
\def\sx{{\hat\sigma_x}}
\def\sy{{\hat\sigma_y}}
\def\sz{{\hat\sigma_z}}
\def\svec{{\hat{\vec\sigma}}}
\def\nvec{{\vec n}}
\def\id{{\hat 1}}
\def\e{{\rm e}}
\def\tr{{\rm Tr}}
\def\H{{\hat H}}
\def\U{{\hat U}}
\def\Udag{{\hat U^\dagger}}
\def\O{{\hat O}}
\def\Odag{{\hat O^\dagger}}
\def\P{{\hat{\cal P}}}
\def\E{{\hat E}}
\def\A{{\hat A}}
\def\Adag{{\hat A^\dagger}}
\def\B{{\hat B}}
\def\Bdag{{\hat B^\dagger}}
\def\L{{\hat L}}
\def\Ldag{{\hat L^\dagger}}
\def\Z{{\hat Z}}

\begin{document}

\title{A simple model of quantum trajectories}

\author{Todd A. Brun\thanks{Email:  tbrun@ias.edu} \\
Institute for Advanced Study, Einstein Drive, \\
Princeton, NJ  08540 }

\maketitle

\begin{abstract}
Quantum trajectory theory, developed largely in the quantum optics
community to describe open quantum systems subjected to continuous
monitoring, has applications in many areas of quantum physics.
In this paper I present a simple model, using two-level quantum
systems (q-bits), to illustrate the essential physics of quantum
trajectories and how different monitoring schemes correspond to different
``unravelings'' of a mixed state master equation.
I also comment briefly on the relationship of the
theory to the Consistent Histories formalism and to spontaneous collapse
models.
\end{abstract}

\section{Introduction}

Over the last ten years the theory of quantum trajectories has been
developed by a wide variety of authors
\cite{Carmichael,Dalibard,Dum,Gardiner,Gisin1,Diosi,Gisin2,Schack}
for a variety of purposes,
including the ability to model continuously monitored open systems
\cite{Carmichael,Dum,Gardiner}, improved numerical calculation
\cite{Dalibard,Schack}, and insight into the problem
of quantum measurement \cite{Gisin1,Diosi,Gisin2}.
However, outside the small communities of quantum optical
theory and quantum foundations, the theory remains little known and
poorly understood.

Indeed, the association with quantum foundations has convinced many
observers that quantum trajectory theory is different from
standard quantum mechanics, and therefore to be regarded with deep
suspicion \cite{Zeh}.  While it is true that stochastic Schr\"odinger
and master equations of the type treated in quantum trajectories
are sometimes postulated in alternative quantum theories
\cite{Pearle,GRW,Percival2,Penrose},
these same types of equations arise quite naturally in describing
quantum systems interacting with environments (open systems)
which are subjected to monitoring by measuring devices.  In these
systems, the stochastic equations arise as {\it effective} evolution
equations, and are in no sense anything other than standard quantum
mechanics (except, perhaps, in the trivial sense of approaching the
limit of continuous measurement).

The fact that this is not widely appreciated, even in fields which might
usefully employ quantum trajectories, is a great pity.  I attribute it
to confusion about exactly what the theory is, and how these equations
arise.

To combat this confusion I present in this paper a simple model for
quantum trajectories, using only two-level quantum systems ({\it quantum bits}
or q-bits).  Because of the recent interest in quantum information
a great deal of attention has been paid to the possible interactions
and measurements of two-level systems; I will draw on this to build my
model and show how effective stochastic evolution equations can arise
due to a system interacting with a continuously monitored environment.

This paper is intended as a tutorial, giving the physical basis of
quantum trajectories without the complications arising from realistic
systems and environments.  As an aid in understanding or teaching the
material, I have included a number of exercises throughout this paper.
For the most part these involve working out derivations which arise in
developing the model, but which are not crucial for following the main
argument.  The solutions to the exercises are included in the appendix.

\subsection{The model}

For simplicity, I will consider the simplest possible quantum system:
a single two-level atom or q-bit.  The environment will also consist of 
q-bits.  While this is far easier to analyze
than the more complicated systems used in most actual quantum trajectory
simulations, it can illustrate almost all of the phenomena which occur
in more realistic cases.  Moreover, the recent interest in
quantum computation and quantum information has spurred a great deal of
work on such systems of q-bits, which we can draw on.

I will assume that the environment bits interact
with the system bit successively for finite, disjoint intervals.
I will further assume that these bits
are all initially in the same state, have no correlations with each other,
and do not interact amongst themselves.  (See figure 1.)  Therefore all
that we need to consider is a succession of interactions between two q-bits.
After each environment bit interacts with the system, we can measure it
in any way we like, and use the information obtained to update our
knowledge of the state.

This model of system and environment may seem excessively abstract,
but in fact a number of experimental systems can be approximated in this
way.  For instance, in experiments with ions in traps, residual molecules
of gas can occasionally pass close to the ion, perturbing its internal
state.  In cavity QED experiments the system might be an electromagnetic
mode inside a high-$Q$ cavity.  If there is never more than a single
photon present, the mode is reasonably approximated as a two-level system.
This mode can be probed by sending Rydberg atoms through the cavity one at
a time in a superposition of two neighboring electronic states,
and measuring the atoms' electronic states upon their emergence.  In this
case, the atoms are serving as both an environment for the cavity mode
and a measurement probe.  Similarly, in an optical microcavity the
external electromagnetic field can be thought of as a series of incoming
waves which are reflected by the cavity walls; photons leaking out from the
cavity can move these modes from the vacuum to the first excited state.
(See figure 2.)

\subsection{Plan of this paper}

In section 2, I describe systems of q-bits, and the possible
two-bit interactions they can undergo.  In section 3, I examine possible
kinds of measurements, including {\it positive operator valued} and
{\it weak} measurements.

In section 4, I present the model of the system and environment.  In
section 5 I show how it is possible to give an effective evolution
for the system alone by tracing out the environment degrees of freedom,
producing a kind of master equation.

In section 6, I consider the effects on the system state of a measurement on
the environment, and deduce stochastic evolution equations for the system
state, conditioned on the random outcome of the environment measurements.

In sections 7--8, I examine different interactions and measurements,
and how they lead to different effective evolution equations for pure
states---a kind of stochastic Schr\"odinger equation.
I show how under some circumstances the system appears to evolve by
large discontinuous jumps, while for other models the state undergoes
a slow diffusion in Hilbert space.  I further show that by averaging over
the different measurement outcomes we recover the master equation evolution
of section 5.

In section 9, I consider measurement schemes which give only partial
information about the system.  These schemes lead to stochastic master
equations rather than Schr\"odinger equations.

Finally, in section 10 I relax the assumption of measurement, and discuss
how quantum trajectories can be used to give insight into the nature of
measurement itself, using the decoherent (Consistent) histories formulation
of quantum mechanics.  I also briefly consider alternative quantum theories
which make use of stochastic equations.  In section 11 I summarize the paper
and draw conclusions.  I present the solution to the exercises in the
appendix.

\section{Two-level systems and their interactions}

The simplest possible quantum mechanical system is a two-level atom or
{\it q-bit}, which has a two-dimensional Hilbert space ${\cal H}_2$.
There are many physical embodiments of such a system:  the spin of a
spin-1/2 particle, the polarization states of a photon, two hyperfine
states of a trapped atom or ion, two neighboring levels of a Rydberg atom,
the presence or absence of a photon in a microcavity, etc.  All of the
above have been proposed in various schemes for quantum information and
quantum computation, and many have been used in actual
experiments.  (For a good general source on quantum computation and
information, see Nielsen and Chuang \cite{NielsenChuang} and
references therein.)

By convention, we choose a particular basis and label its basis states
$\ket0$ and $\ket1$, which we define to be the eigenstates of the Pauli
spin matrix $\sz$ with eigenvalues $-1$ and $+1$, respectively.
We similarly define the other Pauli operators $\sx,\sy$; linear combinations
of these, together with the identity $\id$, are sufficient to produce any
operator on a single q-bit.

The most general pure state of a q-bit is
\begin{equation}
\ket\psi = \alpha\ket0 + \beta\ket1,\ \ |\alpha|^2 + |\beta|^2 = 1 \;.
\label{qbit_pure}
\end{equation}
A global phase may be assigned arbitrarily, so all physically distinct
pure states of a single q-bit form a two-parameter space.
A useful parametrization is in terms of two angular variables
$\theta$ and $\phi$,
\begin{equation}
\ket\psi = \cos(\theta/2)\e^{-i\phi/2} \ket0
  + \sin(\theta/2)\e^{i\phi/2} \ket1 \;,
\label{pure_param}
\end{equation}
where $0 \le \theta \le \pi$ and $0 \le \phi \le 2\pi$.  These two
parameters define a point on the {\it Bloch sphere}.  The north
and south poles of the sphere represent the eigenstates of $\sz$, and
the eigenstates of $\sx$ and $\sy$ lie on the equator.  Orthogonal states
always lie opposite each other on the sphere.

If we allow states to be {\it mixed}, we represent a q-bit by a density
matrix $\rho$; the most general density matrix can be written
\begin{equation}
\rho = p \ket\psi\bra\psi + (1-p) \ket{\bar\psi}\bra{\bar\psi} \;,
\label{orthog_rho}
\end{equation}
where $\ket\psi$ and $\ket{\bar\psi}$ are two orthogonal pure states,
$\bracket\psi{\bar\psi}=0$.  The mixed states of a q-bit
form a three parameter family:
\begin{eqnarray}
\rho &=& \left( {1+r\over2}\cos^2(\theta/2)
  + {1-r\over2}\sin^2(\theta/2) \right) \ket0\bra0 + \nonumber\\
&& \left( {1+r\over2}\sin^2(\theta/2)
  + {1-r\over2}\cos^2(\theta/2) \right) \ket1\bra1 + \nonumber\\
&& r \cos(\theta/2) \sin(\theta/2)
  \left( \e^{i\phi} \ket0\bra1 + \e^{-i\phi} \ket1\bra0 \right) \;,
\label{qbit_mixed}
\end{eqnarray}
where $\theta,\phi$ are the same angular parameters as before and
$0\le r \le 1$.  The limit $r=1$ is the set of pure states, parametrized as
in (\ref{pure_param}), while $r=0$ is the {\it completely mixed state}
$\rho=\id/2$.
Thus we can think of the Bloch sphere as having pure states on its surface
and mixed states in its interior; and the distance $r$ from the center is a
measure of the state's purity.  It is simply related to the parameter
$p$ in (\ref{orthog_rho}):  $p = (1+r)/2$, $1-p = (1-r)/2$.

For {\it two} q-bits, the Hilbert space ${\cal H}_2\otimes{\cal H}_2$
has a tensor-product basis
\begin{eqnarray}
\ket{0}_A\otimes\ket{0}_B &\equiv& \ket{00}_{AB} \;, \nonumber\\
\ket{0}_A\otimes\ket{1}_B &\equiv& \ket{01}_{AB} \;, \nonumber\\
\ket{1}_A\otimes\ket{0}_B &\equiv& \ket{10}_{AB} \;, \nonumber\\
\ket{1}_A\otimes\ket{1}_B &\equiv& \ket{11}_{AB} \;;
\end{eqnarray}
similarly, for $N$ q-bits we can define a basis
$\{ \ket{i_{N-1}i_{N-2}\cdots i_0} \}$, $i_k=0,1$.  A useful labeling of
these $2^N$ basis vectors is by the integers $0\le j < 2^N$ whose
binary expressions are $i_{N-1}\cdots i_0$:
\begin{equation}
\ket{j} \equiv \ket{i_{N-1}\cdots i_0},\ \ 
  j = \sum_{k=0}^{N-1} i_k 2^k \;.
\end{equation}

All states evolve according to the Schr\"odinger equation with some
Hamiltonian $\H(t)$,
\begin{equation}
{d\ket\psi\over dt} = - {i\over\hbar} \H(t) \ket\psi.
\label{schrodinger}
\end{equation}
Over a finite time this is equivalent to applying a unitary operator
$\U$ to the state $\ket\psi$,
\begin{equation}
\U = {\rm T:} \exp \left\{ -{i\over\hbar} \int_{t_0}^{t_f} dt\, \H(t)
  \right\} : \;.
\end{equation}
where ${\rm T:}:$ indicates that the integral should be taken in a
time-ordered sense, with early to late times being composed from right
to left.  For the models I consider in this paper I will treat all time
evolution at the level of unitary transformations rather than explicitly
solving the Schr\"odinger equation, so time can be treated as a discrete
variable
\begin{equation}
\ket{\psi_n} = \U_n \U_{n-1} \cdots \U_1 \ket{\psi_0} \;.
\end{equation}
For a mixed state $\rho$, Schr\"odinger time evolution is
equivalent to $\rho \rightarrow \U \rho \Udag$.  Henceforth, I will
also assume $\hbar=1$.

If the unitary operator $\U_n$ is {\it weak}, that is, close to the
identity, one can always find a Hamiltonian operator $\H_n$ such that
\begin{equation}
\U_n = \exp\left\{ -i \H_n \delta t \right\} \approx 1 - i \H_n \delta t \;.
\end{equation}
Thus, one can easily recover the Schr\"odinger equation from a description
in terms of unitary operators,
\begin{equation}
\delta\ket{\psi_n} = \ket{\psi_n} - \ket{\psi_{n-1}}
  = (\U_n - \id)\ket{\psi_{n-1}} = - i\H_n\ket{\psi_{n-1}} \delta t \;.
\end{equation}

\medskip\noindent
{\bf Exercise 1.}  Show that the most general unitary operator on a single
q-bit can be written
\begin{equation}
\U = \id\cos\varphi + i\nvec\cdot\svec\sin\varphi
  = \exp\{ i\varphi\nvec\cdot\svec \}\;,
\end{equation}
modulo an irrelevant overall phase,
where $\nvec=\{n_x,n_y,n_z\}$ is a three-vector with unit norm and
$\svec = \{\sx,\sy,\sz\}$.  In the Bloch sphere picture this is a rotation
by an angle $\varphi$ about the axis defined by $\nvec$, with three real
parameters (four if one includes the global phase).
\medskip

For two q-bits there is unfortunately no such simple visualization as the
Bloch sphere.  However, to specify any unitary transformation it suffices
to give its effect on a complete set of basis vectors.  I will consider only a
fairly limited set of two-bit transformations in this paper, and no
transformations involving more than two q-bits, but the simple formalism
I derive readily generalizes to higher-dimensional systems.

Let us examine a couple of examples of two-bit transformations.  The
controlled-NOT gate (or CNOT) is widely used in quantum computation; applied
to the tensor-product basis vectors it gives
\begin{eqnarray}
\U_{\rm CNOT} \ket{00} &=& \ket{00} \;, \nonumber\\
\U_{\rm CNOT} \ket{01} &=& \ket{01} \;, \nonumber\\
\U_{\rm CNOT} \ket{10} &=& \ket{11} \;, \nonumber\\
\U_{\rm CNOT} \ket{11} &=& \ket{10} \;,
\end{eqnarray}
or in matrix form
\begin{equation}
\U_{\rm CNOT} = \left( \matrix{ 1 & 0 & 0 & 0 \cr
                        0 & 1 & 0 & 0 \cr
                        0 & 0 & 0 & 1 \cr
                        0 & 0 & 1 & 0 \cr} \right) \;.
\end{equation}
If the first bit is in state $\ket0$ this gate leaves the second bit
unchanged; if the first bit is in state $\ket1$ the second bit is
flipped $\ket0 \leftrightarrow \ket1$.  Hence the name:  whether a NOT
gate is performed on the second bit is {\it controlled} by the first bit.
In terms of single-bit operators this is
$\U_{\rm CNOT} = \ket0\bra0 \otimes \id
+ \ket1\bra1 \otimes \sx$.

Another important gate in quantum computation is the SWAP; applied
to the tensor-product basis vectors it gives
\begin{eqnarray}
\U_{\rm SWAP} \ket{00} &=& \ket{00} \;, \nonumber\\
\U_{\rm SWAP} \ket{01} &=& \ket{10} \;, \nonumber\\
\U_{\rm SWAP} \ket{10} &=& \ket{01} \;, \nonumber\\
\U_{\rm SWAP} \ket{11} &=& \ket{11} \;,
\end{eqnarray}
which in matrix form is just
\begin{equation}
\U_{\rm SWAP} = \left( \matrix{ 1 & 0 & 0 & 0 \cr
                        0 & 0 & 1 & 0 \cr
                        0 & 1 & 0 & 1 \cr
                        0 & 0 & 0 & 1 \cr} \right) \;.
\end{equation}
As the name suggests, the SWAP gate just exchanges the states of the
two bits:  $\U_{\rm SWAP}(\ket\psi\otimes\ket\phi)
= \ket\phi\otimes\ket\psi$.

CNOT and SWAP are examples of two-bit {\it quantum gates}.  Such gates are
of tremendous importance in the theory of quantum computation.  These two
gates in particular have an additional useful property.  Note that
the operator $\U_{\rm CNOT} = \Udag_{\rm CNOT}$, i.e., it is both
unitary and Hermitian.  This is also true of $\U_{\rm SWAP}$.  This is only
possible if all of the operator's eigenvalues are $1$ or $-1$.  This
also means that these operators are their own inverses,
$\U^2_{\rm CNOT} = \U^2_{\rm SWAP} = \id$.  (Among single-bit operators,
the Pauli matrices ${\hat\sigma}_{x,y,z}$ also have this property,
as does any operator ${\vec n}\cdot\svec$.)

This property of being both unitary and Hermitian makes it possible to
define one-parameter families of two-bit unitary transformations:
\begin{eqnarray}
\U_{\rm CNOT}(\theta) &=&
  \exp\left\{ - i \theta \U_{\rm CNOT} \right\} \nonumber\\
&=& \id \cos\theta - i \U_{\rm CNOT} \sin\theta \;,
\end{eqnarray}
and similarly 
\begin{eqnarray}
\U_{\rm SWAP}(\theta) &=&
  \exp\left\{ - i \theta \U_{\rm SWAP} \right\} \nonumber\\
&=& \id \cos\theta - i \U_{\rm SWAP} \sin\theta \;.
\end{eqnarray}

\medskip\noindent
{\bf Exercise 2.}  Show that any operator $\U$ which is both unitary and
Hermitian satisfies $\exp(-i\theta\U) = \id\cos\theta - i \U\sin\theta$.
\medskip

These families range from the
identity $\id$ for $\theta=0$ to the full CNOT (or SWAP) gate for
$\theta=\pi/2$, up to a global phase.
For small values $\theta \ll 1$, this is a weak
interaction, which leaves the state only slightly altered.

These families do have an undesirable characteristic, however.  Suppose
we apply $\U_{\rm CNOT}(\theta)$ to a two-bit state of the form
$(\alpha\ket0 + \beta\ket1)\otimes\ket\psi$.  The new state is
\begin{eqnarray}
\U_{\rm CNOT}(\theta) (\alpha\ket0 + \beta\ket1)\otimes\ket\psi
  &=& (\alpha\e^{-i\theta}\ket0 + \beta\cos\theta\ket1) \otimes \ket\psi
  \nonumber\\
&&  - i \beta\sin\theta\ket1\otimes(\sx\ket\psi) \;.
\end{eqnarray}
This relative phase $\e^{i\theta}$ between $\ket0$ and $\ket1$ is a
complication in the calculations that will follow.  To avoid this problem,
instead of $\U_{\rm CNOT}(\theta)$ and $\U_{\rm SWAP}(\theta)$ we
will use interactions of the form
$\Z_A(\theta)\U_{\rm CNOT}(\theta)$, where
$\Z_A(\theta) = (\exp(-i\sz\theta/2)\otimes\id)$,
and similarly for $\U_{\rm SWAP}$.  The single-bit rotation exactly undoes the 
extra relative phase produced by $\U_{\rm CNOT}(\theta)$, while changing
nothing else.

Before leaving the topic of two-bit gates, I should point out
another representation which is sometimes useful.  Any operator on two
q-bits can formally be written
\begin{equation}
\O = \sum_{i,j=0}^3 R_{ij} \hat\sigma_i \otimes \hat\sigma_j,
\end{equation}
where $\hat\sigma_0\equiv\id$, $\hat\sigma_{1,2,3}\equiv\hat\sigma_{x,y,z}$.
The matrix elements can be calculated using the identity
\begin{equation}
R_{ij} = \tr \left\{ \O (\sigma_i\otimes\sigma_j) \right\} \;.
\end{equation}
If $\O=\Odag$ is Hermitian, the matrix elements $R_{ij}$ will be real.

If we write the CNOT in this form,
it gives matrix elements $R_{ij}$
\begin{equation}
{\bf R}_{\rm CNOT} = \left( \matrix{ 1/2 & 1/2 & 0 & 0 \cr
                          0 & 0 & 0 & 0 \cr
                          0 & 0 & 0 & 0 \cr
                          -1/2 & 1/2 & 0 & 0 \cr} \right) \;.
\end{equation}
Similarly, if we write the SWAP gate this way it has
matrix elements
\begin{equation}
{\bf R}_{\rm SWAP} = \left( \matrix{ 1/2 & 0 & 0 & 0 \cr
                          0 & 1/2 & 0 & 0 \cr
                          0 & 0 & 1/2 & 0 \cr
                          0 & 0 & 0 & 1/2 \cr} \right) \;.
\end{equation}

\section{Strong and weak measurements}

\subsection{Projective measurements}

In the standard description of quantum mechanics, observables are identified
with Hermitian operators $\O=\Odag$.  A measurement returns an eigenvalue
$o_n$ of $\O$ and leaves the system in the corresponding eigenstate
$\ket{\phi_n}$, $\O\ket{\phi_n}=o_n\ket{\phi_n}$, with probability
$p_n=|\bracket{\phi_n}{\psi}|^2$.  If a particular eigenvalue is degenerate,
one instead uses the projector $\P_n$ onto the eigenspace with eigenvalue
$o_n$; the probability of the measurement outcome is then
$p_n = \bra\psi\P_n\ket\psi$ and the system is left in the state
$\P_n\ket\psi/\sqrt{p_n}$.  For a mixed state $\rho$ the probability of
outcome $n$ is $p_n = \tr\{\P_n\rho\}$ and the state after the measurement
is $\P_n\rho\P_n/p_n$.

Because two observables with the same eigenspaces are completely equivalent
to each other (as far as measurement probabilities and outcomes
are concerned), we will not worry about the exact choice
of Hermitian operator $\O$; instead, we will
choose a complete set of orthogonal projections $\{\P_n\}$ which
represent the possible measurement outcomes.  These satisfy
\begin{equation}
\P_n \P_{n'} = \P_n \delta_{nn'} \;, \ \ \sum_n \P_n = \id \;.
\label{orthogonal_decomp}
\end{equation}
A set of projection operators which obey (\ref{orthogonal_decomp}) is
often referred to as an {\it orthogonal decomposition of the identity}.  For
a single q-bit, the only nontrivial measurements have exactly two outcomes,
which we label $+$ and $-$, with probabilities $p_+$ and $p_-$ and
associated projectors of the form
\begin{equation}
\P_\pm = { \id \pm \nvec\cdot\svec \over 2 }
  = \ket{\psi_\pm}\bra{\psi_\pm} \;;
\end{equation}
this is equivalent to choosing an axis $\vec n$
on the Bloch sphere and projecting
the state onto one of the two opposite points.
All such operators are projections onto pure states.
The two projectors sum to the identity operator, $\P_++\P_-=\id$.
The average information obtained from a projective measurement on a q-bit is
just the Shannon entropy for the two measurement outcomes,
\begin{equation}
S_{\rm meas} = - p_+ \log_2 p_+ - p_- \log_2 p_- \;.
\label{shannon}
\end{equation}
The maximum information gain is precisely one bit, when $p_+=p_-=1/2$,
and the minimum is zero bits when either $p_+$ or $p_-$ is 0.
After the measurement, the state is left in an eigenstate of $\P_\pm$, so
repeating the measurement will result in the same outcome.  This repeatability
is one of the most important features of projective measurements.

\subsection{Positive operator valued and weak measurements}

While projective measurements are the most familiar from introductory
quantum mechanics, there is a more general notion of measurement, the
{\it positive operator valued measurement} (POVM) \cite{Peres}.
Instead of giving a set of projectors which sum to the identity, we
give a set of {\it positive operators} $\E_n$ which sum
to the identity:
\begin{equation}
\sum_n \E_n = \id \;.
\end{equation}
The probability of outcome $n$ is $p_n = \bra\psi\E_n\ket\psi$, or for
a mixed state $p_n = \tr\{\E_n\rho\}$.  Unlike a projective measurement,
knowing the operators $\E_n$ is {\it not} sufficient to determine the
state of the system after measurement.  One must further know a set of
operators $\A_{nk}$ such that
\begin{equation}
\E_n = \sum_k \Adag_{nk}\A_{nk} \;.
\end{equation}
After measurement outcome $n$ the state is
\begin{equation}
\rho \rightarrow \rho' = (1/p_n) \sum_k \A_{nk}\rho \Adag_{nk} \;.
\end{equation}
This measurement will not preserve the purity of states, in general, unless
there is only a single $\A_{nk}$ for each $\E_n$ (that is,
$\E_n = \Adag_n\A_n$), in which case $\ket{\psi'} = \A_n\ket\psi/\sqrt{p_n}$.

Because the positive operators $\E_n$ need not be projectors, one is not
limited to only two possible outcomes; indeed, there can be an unlimited
number of possible outcomes.  However, if a POVM is repeated, the same
result will not necessarily be obtained the second time.  Projective
measurements are clearly a special case of POVMs in which results {\it are}
repeatable.  Most actual experiments do {\it not} correspond to projective
measurements, but are described by some more general POVM.

Conversely, it is easy to show that any POVM can be performed by allowing
the system which is to be measured to interact with an
additional system, or {\it ancilla}, and then doing a projective measurement
on the ancilla.  In this viewpoint, only some of the information obtained
comes from the system; part can also come from the ancilla, which injects
extra randomness into the result.  This mathematical identification
should not be pushed too far, however; in practice, trying to interpret a
POVM as a projective measurement on an ancilla often adds complexity without
improving understanding.

A particularly interesting kind of POVM for the present purpose is the
{\it weak measurement} \cite{Vaidman}.
This is a measurement which gives very
little information about the system on average, but also disturbs the
state very little.

Loosely speaking, there are two ways a measurement can be
considered weak.  Suppose we have a q-bit in a state of form
(\ref{qbit_pure}), and we perform a POVM with the following two operators:
\begin{eqnarray}
\E_0 &\equiv& \ket0\bra0 + (1-\epsilon)\ket1\bra1
  = \A_0^2 \;, \nonumber\\ 
\E_1 &\equiv& \epsilon \ket1\bra1 = \A_1^2 \;, \nonumber\\
\A_0 &\equiv& \ket0\bra0 + \sqrt{1-\epsilon}\ket1\bra1
  \;, \nonumber\\ 
\A_1 &\equiv& \sqrt\epsilon \ket1\bra1 \;,
\label{weak1}
\end{eqnarray}
where $\epsilon \ll 1$.  Clearly $\E_0$ and $\E_1$ are positive and
$\E_0+\E_1=\id$, so this is a POVM.
The probability $p_0 = \bra\psi\E_0\ket\psi = 1-\epsilon|\beta|^2$
of outcome 0 is close to 1, while
$p_1 = \bra\psi\E_1\ket\psi = \epsilon|\beta|^2$ is
very unlikely.  Thus, most such measurements will give outcome 0, and very
little information is obtained about the system.

\medskip\noindent
{\bf Exercise 3.}  Show that the average information
gain from this measurement (in bits) is
\begin{equation}
S_{\rm meas} \approx  \epsilon|\beta|^2(1/\ln 2 - \log_2 \epsilon|\beta|^2)
  \ll 1 \;.
\end{equation}
\medskip

The state changes very slightly after a measurement outcome 0,
with $\ket0$ becoming slightly more likely relative to $\ket1$.
The new state is
\begin{equation}
\ket{\psi_0} = \A_0\ket\psi/\sqrt{p_0}
  = (\alpha\ket0 + \beta \sqrt{1-\epsilon}\ket1)/\sqrt{p_0} \;.
\end{equation}

However, after an outcome of 1, the state can change dramatically:  a
measurement outcome of 1 leaves the q-bit in the state $\ket1$.  We see
that this type of measurement is weak in that it {\it usually} gives little
information, and has little effect on the state; but on rare occasions it
can give a great deal of information and have a large effect.

By contrast, consider the following positive operators:
\begin{eqnarray}
\E_0 &\equiv& \left( {1+\epsilon \over 2} \right) \ket0\bra0
  + \left( {1-\epsilon \over 2} \right) \ket1\bra1
  = \A_0^2 \;,\nonumber\\ 
\E_1 &\equiv& \left( {1-\epsilon \over 2} \right) \ket0\bra0
  + \left( {1+\epsilon \over 2} \right) \ket1\bra1
  = \A_1^2 \;, \nonumber\\
\A_0 &\equiv& \sqrt{1+\epsilon \over 2} \ket0\bra0
  + \sqrt{1-\epsilon \over 2} \ket1\bra1 \;, \nonumber\\
\A_1 &\equiv& \sqrt{1-\epsilon \over 2} \ket0\bra0
  + \sqrt{1+\epsilon \over 2} \ket1\bra1 \;.
\label{weak2}
\end{eqnarray}
These operators also constitute a POVM.  Both $\E_0$ and $\E_1$ are close
to $\id/2$, and so are almost equally likely for all states,
$p_0 \approx p_1 \approx 1/2$; the information acquired is approximately
one bit.  But the state of the system is almost unchanged for {\it both}
outcomes, with the new state being $\ket{\psi_j} = \A_j\ket\psi/\sqrt{p_j}$,
\begin{eqnarray}
\ket{\psi_0} &=& {1\over\sqrt{1+\epsilon(|\alpha|^2-|\beta|^2)}}
  \left( \alpha\sqrt{1+\epsilon}\ket0
  + \beta\sqrt{1-\epsilon}\ket1 \right) \nonumber\\
&\approx& \alpha(1+\epsilon|\beta|^2)\ket0
  + \beta(1-\epsilon|\alpha|^2)\ket1 \;, \nonumber\\
\ket{\psi_1} &=& {1\over\sqrt{1-\epsilon(|\alpha|^2-|\beta|^2)}}
  \left( \alpha\sqrt{1-\epsilon}\ket0
  + \beta\sqrt{1+\epsilon}\ket1 \right) \nonumber\\
&\approx& \alpha(1-\epsilon|\beta|^2)\ket0
  + \beta(1+\epsilon|\alpha|^2)\ket1 \;,
\end{eqnarray}
for outcomes 0 and 1, respectively.  For this type of weak measurement,
the measurement outcome is almost random, but does include a tiny amount
of information about the state.  By performing repeated weak measurements
and examining the statistics of the results, one can in effect perform
a strong measurement; for the particular case considered here, the
state will tend to drift towards either $\ket0$ or $\ket1$ with
overall probabilities $|\alpha|^2$ and $|\beta|^2$.

We will see below specific instances of how a POVM can arise by the
interaction of the measured system with an ancilla,
and the subsequent projective measurement of the
ancilla.  Quantum trajectories are
closely related to such indirect measurement schemes.

\section{System and environment}

In discussing quantum evolution it is usually assumed that the quantum
system is very well isolated from the rest of the world.  This is a
useful idealization, but it is rarely realized in practice, even
in the laboratory.  In fact, most systems interact at least
weakly with external degrees of freedom \cite{Zurek,Joos}.

One way of taking this into account is to include
a model of these external degrees
of freedom in our description.  Let us assume that in addition
to the system in state $\ket\psi$ in Hilbert space ${\cal H}_S$
with Hamiltonian $\H_S$, there is an external {\it environment}
in state $\ket{E}\in{\cal H}_E$ with Hamiltonian $\H_E$.
The joint Hilbert space of the system and environment is the tensor product
space ${\cal H}_S\otimes{\cal H}_E$, and the full Hamiltonian
can be written
\begin{equation}
\H = \H_S\otimes\id_E + \id_S\otimes\H_E + \H_I \;,
\end{equation}
where $\H_I$ is a Hamiltonian giving the interaction between the two.
The system and environment evolve according to the Schr\"odinger
equation (\ref{schrodinger}) with this full Hamiltonian.  Because of
the presence of the $\H_I$ term, in general even initial product states
$\ket\psi\otimes\ket{E}$ will evolve into states which cannot be
written as product states:
\begin{equation}
\ket\psi\otimes\ket{E} \rightarrow
  \ket\Psi = \sum_j \ket{\psi_j}\otimes\ket{E_j} \;.
\label{entangled}
\end{equation}
Such states are called {\it entangled}.  If the system and environment
are entangled, it is no longer possible to attribute a unique pure state
to the system (or environment) alone.  The best that can be done is to
describe the system as being in a mixed state $\rho_{\rm sys}$, which
is the partial trace of the joint state $\ket\Psi$ over the environment
degrees of freedom:
\begin{eqnarray}
\rho_{\rm sys} &=& \tr_{\rm env} \left\{ \ket\Psi\bra\Psi \right\} \nonumber\\
&=& \sum_{j,j',k} \ket{\psi_j}\bra{\psi_{j'}}
  \bracket{e_k}{E_j}\bracket{E_{j'}}{e_k} \nonumber\\
&=& \sum_{j,j'} \ket{\psi_j}\bra{\psi_{j'}}
  \bracket{E_{j'}}{E_j} \;,
\end{eqnarray}
where the $\{\ket{e_k}\}$ are some complete set of orthonormal basis vectors
on the environment Hilbert space ${\cal H}_E$.

The entanglement of a state can be quantified.  The most widely used
measure for pure states is the {\it entropy of entanglement}.
For a system and environment in a joint pure state, this is
\begin{equation}
S_E = - \tr\{\rho_S \log_2 \rho_S \} \;,
\end{equation}
which is the {\it von~Neumann entropy} of the reduced density operator
$\rho_S$.  This entropy is maximized by the density operator
$\rho_S = \id_S/\tr_{\rm sys}\{\id_S\}$; such a density operator
is called {\it maximally mixed},
and the joint pure state which gives rise to it is called
{\it maximally entangled}.

\medskip\noindent
{\bf Exercise 4.}  Show that the von~Neumann entropy is maximized for
$\rho = \id/\tr\{\id\}.$
\medskip

Systems which interact with their environments are said to be {\it open}.
Most real physical environments are extremely complicated, and the
interactions between systems and environments are often poorly understood.
In analyzing open systems, one often makes the approximation of assuming
a simple, analytically solvable form for the environment degrees of
freedom.

For the models of this paper, I will assume that both the system and
the environment consist solely of q-bits.  I will also assume a simple
form of interaction, namely that the system q-bit interacts with one
environment q-bit at a time, and that after interacting they never come
into contact again; and that the environment q-bits have no Hamiltonian
of their own, $\H_E=0$.  This environment may seem
ridiculously oversimplified, but in fact it suffices to demonstrate most
of the physics exhibited by much more realistic descriptions.  And in
fact, the description in terms of q-bits is not a bad schematic picture
of many environments at low energies.  I summarize one such
possible environment in Figure 2.

For this type of model, the Hilbert space of the system is
${\cal H}_S = {\cal H}_2$ and the Hilbert space of the environment
is ${\cal H}_E = {\cal H}_2 \otimes {\cal H}_2 \otimes \cdots$.
I will in general assume that all the environment q-bits start in some
pure initial state, usually $\ket0$; however, in later elaborations of
the model I will relax this assumption to include
both other pure states and mixed-state environments,
which resemble heat baths of finite temperature.

\section{The master equation}

A system interacting with an environment usually cannot
be described in terms of a pure state. Similarly, it is not usually possible
to give a simple time-evolution for the system alone, without reference
to the state of the environment.  However, under certain
circumstances one can find an effective evolution equation
for the system alone, if the interaction between the system and
environment has a simple form and the initial state of the
environment has certain properties.

In cases where such an effective evolution equation exists,
it will not be of the form (\ref{schrodinger}).
Instead, the evolution equation will be {\it completely positive},
which is a weaker condition than unitarity.  It must
take density matrices to density matrices, but not necessarily pure states
to pure states.  To be tractable, such an equation must also generally be
{\it Markovian}:  it must give the evolution of the density operator
solely in terms of its state at the present time, with no retarded terms.
(While it is possible to find retarded descriptions for more general
non-Markovian open systems, they tend to be difficult to solve, even
numerically.)  The most general such Markovian equation \cite{Lindblad}
is a master equation in {\it Lindblad} form
\begin{equation}
{d\rho\over dt} = - {i\over\hbar}[\H_S,\rho] +
  \sum_k \left[ \L_k \rho \Ldag_k - (1/2) \Ldag_k\L_k\rho
  - (1/2)\rho\Ldag_k\L_k \right] \;,
\label{Lindblad}
\end{equation}
where $\H_S$ is the Hamiltonian for the system alone and the
{\it Lindblad operators} $\{\L_k\}$ give the effects of interaction
with the environment.

What about the kind of discrete time evolution we've been assuming?
Suppose that a system and environment in an initial product state
undergo some unitary transformation $\U$:
\begin{equation}
\ket\Psi = \ket\psi\otimes\ket{E}
  \rightarrow \ket{\Psi'} = \U \left( \ket\psi\otimes\ket{E} \right) \;.
\end{equation}
How does the state of the system alone change?  We can write any operator
on ${\cal H}_S\otimes{\cal H}_E$ as a sum of product
operators
\begin{equation}
\U = \sum_j \A_j \otimes \B_j \;.
\end{equation}
The state of the system after the unitary transformation is
\begin{eqnarray}
\rho'_S &=& \tr_{\rm env} \{ \U \ket\Psi\bra\Psi \Udag \} \nonumber\\
&=&  \sum_{j,j'} \tr_{\rm env} \left\{
  (\A_j\ket\psi\bra\psi\Adag_{j'})\otimes
  (\B_j\ket{E}\bra{E}\Bdag_{j'}) \right\} \nonumber\\
&=&  \sum_{j,j'}
  \A_j\ket\psi\bra\psi\Adag_{j'}
  \bra{E}\Bdag_{j'}\B_j\ket{E} \;.
\label{rhosys}
\end{eqnarray}
The self-adjoint matrix $M_{jj'} \equiv \bra{E}\Bdag_{j'}\B_j\ket{E}$
has a set of orthonormal eigenvectors $\vec\mu_k = \{\mu_{kj}\}$
with real eigenvalues $\lambda_k$ such that
\begin{equation}
\sum_{j'} M_{jj'}\mu_{kj'} = \lambda_k \mu_{kj}\ \ 
  \Rightarrow\ \ M_{jj'} = \sum_k \lambda_k \mu_{kj}\mu^*_{kj'} \;.
\label{eigenvecs}
\end{equation}
We define new operators $\O_k$ by
\begin{equation}
\O_k \equiv \sqrt{\lambda_k} \sum_j \mu_{kj} \A_j \;.
\label{eigen_operators}
\end{equation}

\medskip\noindent
{\bf Exercise 5.}  Show that the unitarity of $\U$ implies
\begin{equation}
\sum_k \Odag_k\O_k = \id_S \;.
\end{equation}
\medskip

The above expression (\ref{rhosys}) then simplifies to
\begin{equation}
\rho'_S = \tr_{\rm env} \{ \U \ket\Psi\bra\Psi \Udag \}
  = \sum_k \O_k \ket\psi\bra\psi \Odag_k \;.
\end{equation}
This expression is quite like the outcome of a POVM, but without determining
a particular measurement result; the system is left in a mixture of all
possible outcomes.

Suppose now that the system interacts with a second environment q-bit in
the same way, with the second bit in the same initial state $\ket{E}$,
and that the Hilbert space of the of two environment bits is
traced out.  The system state will become
\begin{equation}
\rho''_S =
  \sum_{k_1,k_2} \O_{k_2} \O_{k_1} \ket\psi\bra\psi \Odag_{k_1} \Odag_{k_2} \;.
\end{equation}
After interacting successively with $n$ environment bits the system state is
\begin{equation}
\rho^{(n)}_S =
  \sum_{k_1,\ldots,k_n} \O_{k_n} \cdots \O_{k_1}
  \ket\psi\bra\psi \Odag_{k_1} \cdots \Odag_{k_n} \;.
\label{discrete_master}
\end{equation}
This evolution is a type of discrete master equation.  We see clearly that
an initially pure state will in general evolve into a mixed state.  Depending
on the $\O_k$, it may or may not converge to a unique final state.

Let's consider the special case when the unitary operator $\U$ is close
to the identity,
\begin{equation}
\U = \exp\left\{ - i \epsilon \sum_j \A_j\otimes\B_j \right\} \;,
\end{equation}
where $\epsilon\ll 1$, $\tr\{\Adag_j\A_j\} \sim \tr\{\Bdag_j\B_j\} \sim O(1)$,
and the $\A_j$ and $\B_j$ are Hermitian.  Expanding
to second order in $\epsilon$, we see that the new density matrix for the
system is
\begin{eqnarray}
\rho'_S &=& \tr_{\rm env} \{ \U \ket\Psi\bra\Psi \Udag \} \nonumber\\
&\approx& \ket\psi\bra\psi - 
  i \epsilon \sum_j [\A_j, \ket\psi\bra\psi] \bra{E}\B_j\ket{E} \nonumber\\
&& + {\epsilon^2\over2} \sum_{jj'} \bra{E}\B_{j'}\B_j\ket{E} \nonumber\\
&& \times
  \left( 2 \A_j \ket\psi\bra\psi \A_{j'} - \A_{j'}\A_j \ket\psi\bra\psi
  - \ket\psi\bra\psi \A_{j'}\A_j  \right) \;.
\label{infinitesimal_rho}
\end{eqnarray}
Let's make the simplifying assumption that the first-order term vanishes:
\begin{equation}
\sum_j \A_j \bra{E}\B_j\ket{E} = 0 \;.
\end{equation}
Later in the paper we supplement $\U_{\rm CNOT}(\theta)$ and
$\U_{\rm SWAP}(\theta)$ with one-bit gates on the system to remove this
first-order term, which simplifies the evolution.

Let the duration of the interaction be $\delta t$.
We can define a matrix $M_{jj'}$
\begin{equation}
M_{jj'} = \bra{E}\B_{j'}\B_j\ket{E} \;,
\end{equation}
with orthonormal eigenvectors $\vec\mu_k = \{\mu_{kj}\}$ and real eigenvalues
$\lambda_k$ just as in (\ref{eigenvecs}), and define operators
\begin{equation}
\L_k = \sqrt{\epsilon^2\lambda_k\over\delta t} \sum_j \mu_{kj}\A_j \;,
\end{equation}
similarly to (\ref{eigen_operators}) above.
In terms of these operators, equation (\ref{infinitesimal_rho})
can be re-written
\begin{equation}
{{\rho'-\rho}\over \delta t} = \sum_k \left[
  \L_k \rho \Ldag_k - (1/2) \Ldag_k\L_k\rho - (1/2)\rho\Ldag_k\L_k
  \right] \;,
\label{approx_master}
\end{equation}
which has exactly the same form as (\ref{Lindblad}) with a vanishing
Hamiltonian.  If the first order term of (\ref{infinitesimal_rho})
does not vanish, equation (\ref{approx_master}) can pick up an effective
system Hamiltonian, though some care must be exercised about the size
of the various terms and their order in $\delta t$.  But for the simple
case we are considering, we can recover
a continuous evolution equation as a limit of successive brief,
infinitesimal transformations, just as we recovered the Schr\"odinger
equation from a succession of weak unitary transformations.

If we wanted to express the operators $\O_k$ of equation
(\ref{eigen_operators}) in terms of the $\L_k$ for
this infinitesimal transformation, then to second order in $\epsilon$
we would find $\O_k \equiv \L_k\sqrt{\delta t}$, plus one additional
operator $\O_0 \equiv \id - (1/2)\sum_k \Ldag_k\L_k \delta t$.
Substituting these operators in (\ref{discrete_master}) reproduces
(\ref{approx_master}).

\section{Environmental measurements and \\ conditional evolution}

Suppose that the system q-bit is in initial
state $\alpha\ket0+\beta\ket1$ and it
interacts with an environment q-bit in state $\ket0$, so that
a CNOT is performed from the system onto
the environment bit.  The new joint state of the two q-bits is
\begin{equation}
\ket{\Psi'} = \alpha\ket0_S\otimes\ket0_E
  + \beta\ket1_S\otimes\ket1_E \;.
\end{equation}
If we trace out the environment bit the system is in the mixed state
\begin{equation}
\rho_S = |\alpha|^2 \ket0\bra0 + |\beta|^2 \ket1\bra1 \;.
\end{equation}
Now suppose that we make a projective measurement on the environment bit
in the usual $z$ basis.  With probability $p_0 = |\alpha|^2$
($p_1 = |\beta|^2$) we find the bit in state $\ket0$ ($\ket1$), in which
case the system bit will {\it also} be projected into state
$\ket0$ ($\ket1$).  If the system interacts in the same way with more
environment bits, which are also subsequently measured in the $z$ basis,
the same result will be obtained each time with probability 1.  This
scheme acts just like a projective measurement on the system.

This need not be the case in general.  Suppose that instead of a CNOT,
a SWAP gate is performed on the two q-bits.  In this case the new joint
state after the interaction is
\begin{equation}
\ket{\Psi'} = \ket0_S \otimes
  \left( \alpha\ket0_E + \beta\ket1_E \right) \;.
\end{equation}
A subsequent $z$ measurement on the environment yields 0 or 1 with the
same probabilities as before, but now in both cases the system is left
in state $\ket0$.  Further interactions and measurements will always
produce the result 0.  Clearly, the choice of measurement on the environment
bit makes no difference to the state of the system---it will be $\ket0$
for any measurement result, or for no measurement at all.  This is rather
like spontaneous decay of a system which is initially in a superposition
of the ground and excited states.  If we think of $\ket1$ as an excited
state, the system passes on an excitation to the environment with probability
$|\beta|^2$ and is afterwards left in the ground state.
Note that $\U_{\rm CNOT}$ can produce entanglement between two
initially unentangled q-bits, while $\U_{\rm SWAP}$ cannot.

In the case of the CNOT interaction, what happens if we vary our choice of
measurement?  Suppose that instead of measuring the environment bit in
the $z$ basis, we measure it in the $x$ basis
$\ket{x_\pm} = (\ket0 \pm \ket1)/\sqrt{2}$?  In terms of this basis we can
rewrite the joint state after the interaction as
\begin{equation}
\ket{\Psi'} = {1\over\sqrt2} (\alpha\ket0 + \beta\ket1)\otimes\ket{x_+}
  + {1\over\sqrt2} (\alpha\ket0 - \beta\ket1)\otimes\ket{x_-} \;.
\end{equation}
The $+$ and $-$ results are equally likely; the $+$ result leaves the
system state unchanged, while the $-$ flips the relative sign between the
$\ket0$ and $\ket1$ terms.  Aside from knowing whether a flip has occurred,
this measurement result yields no information about the system state;
if this state is initially unknown, it will remain unknown after the
interaction and measurement.  By changing the choice of measurement one
can go from learning the exact state of the system to learning nothing
about it whatsoever.

One important thing to note is that, whatever measurement is performed and
whatever outcome is obtained, the system and environment bit are afterwards
left in a product state.  No entanglement remains.  Thus, the environment
bit can be further manipulated in any way, or discarded completely, without
any further effect on the state of the system.  Because of this, it is
unnecessary to keep track of the environment's state to describe
the system.  This is a somewhat unrealistic approximation,
arising from the assumption of projective measurements---in
real experiments, some residual entanglement with the environment almost
certainly exists which the experimenter is unable to measure or control.  It
does, however, enormously simplify the description of the system's evolution.
I will later relax this assumption, but for now I make use of it.

\section{Stochastic Schr\"odinger equations with \\ jumps}

Far more complicated dynamics result if we replace the strong two-bit
gates of the previous section with weak interactions, such as
$\U_{\rm CNOT}(\theta)$ or $\U_{\rm SWAP}(\theta)$ for $\theta \ll 1$.
In this case the measurements on the environment bits correspond to
{\it weak measurements} on the system, and the system state evolves
unpredictably according to the outcome of the measurements.
Weak interaction with the environment is the norm for most laboratory
systems currently studied; indeed, it is only the weakness of this
interaction which justifies the division into system and environment
in the first place, and enables one to treat the system as isolated to
first approximation, with the effects of the environment as
perturbations.

This problem is simple enough that we can analyze it for a general
weak interaction.  Suppose that the system and environment q-bits are
initially in the state $\ket\Psi = \ket\psi_S\otimes\ket0_E$.
Let the interaction be a two q-bit unitary transformation of the form
\begin{equation}
\U = \exp\left\{ - i \theta \sum_j \A_j\otimes\B_j \right\} \;,
\end{equation}
where $\theta\ll1$,
and $\tr\{\A_j\} \sim \tr\{\B_j\} \sim O(1)$, so we
can expand the exponential in powers of $\theta$.  The new state of
the system and environment is
\begin{eqnarray}
\U\ket\Psi &=& \ket\psi_S\otimes\ket0_E
  - i \theta \sum_j \A_j\ket\psi_S\otimes\B_j\ket0_E
  \nonumber\\
&& - {\theta^2\over2}\sum_{jj'} \A_{j'}\A_j\ket\psi_S \otimes
  \B_{j'}\B_j\ket0_E + O(\theta^3) \;.
\end{eqnarray}

Just as in section 5, we define the matrix
$M_{jj'} = \bra0\B_{j'}\B_j\ket0$ with eigenvectors ${\vec\mu}_k$ and
real eigenvalues $\lambda_k$, and use it to find a new set of operators
\begin{equation}
\L_k = \sqrt{\theta^2\lambda_k\over\delta t} \sum_j \mu_{kj} \A_j \;,
\end{equation}
where $\delta t$ is the time between interactions with successive
environment q-bits.
Since we can decompose the matrix $M_{jj'}$ in terms of two vectors
\begin{equation}
M_{jj'} = \bra0\B_{j'}\ket0\bra0\B_j\ket0
  + \bra0\B_{j'}\ket1\bra1\B_j\ket0 = u_{j'}u_{j} + v_{j'}^*v_j \;,
\end{equation}
it has at most two nonvanishing eigenvalues.
We can make the additional simplifying assumption, as in section 5, that
\begin{equation}
\sum_j \A_j \bra0\B_j\ket0 = 0 \;.
\label{first_order}
\end{equation}
This leaves only a single nonvanishing Lindblad operator,
\begin{equation}
\L = \sqrt{\theta^2\over\delta t} \sum_j \A_j \bra1\B_j\ket0 \;,
\end{equation}
and no effective Hamiltonian term.
In terms of $\L$, the joint state of the system and environment is
\begin{equation}
\U\ket\Psi = (\id - (1/2)\Ldag\L \delta t) \ket\psi_S\otimes\ket0_E
  - i \sqrt{\delta t} \L\ket\psi_S\otimes\ket1_E
  + O(\theta^3) \;.
\end{equation}

The time dependence of $\L \sim \theta/\sqrt{\delta t}$ may seem a bit
strange at first.  It is chosen so that the master equation
(\ref{approx_master}) properly approximates a time derivative.
This does, however, imply a somewhat odd scaling.  The strength of
the interaction between the system and environment q-bits can be
parametrized by $\theta/\delta t$.  This implies that if the two bits
interacted for only half as long, the Lindblad operators would have
to be redefined, $\L \rightarrow \L/\sqrt{2}$.  In the limit as
$\delta t \rightarrow 0$, the right-hand-side of (\ref{approx_master})
would vanish, and the system would not evolve.  This is actually
true, and is an example of the so-called Quantum Zeno Effect \cite{Misra}.
Most such master equations have some microscopic timescale built
in---for instance, a characteristic collision time---and are not
really valid at shorter timescales.  These equations are used to
describe the system at times long compared to this short scale, which
in the case of our model would be after the system had interacted with
a number of environment q-bits.

Once the interaction has taken place,
we measure the environment q-bit in the $\{\ket0,\ket1\}$ basis.
Using the assumption (\ref{first_order}),
the probabilities $p_{0,1}$ of the outcomes are
\begin{eqnarray}
p_0 &=& 1 - \theta^2 \sum_{jj'} \bra\psi\A_{j'}\A_j\ket\psi
  \bra0\B_{j'}\ket1\bra1\B_j\ket0 \nonumber\\
&=& 1 - \bra\psi \Ldag\L \ket\psi \delta t \nonumber\\
&=& 1 - p_1 \;.
\end{eqnarray}
The change in the {\it system} state after a measurement outcome 0
on the {\it environment} is
\begin{equation}
\delta\ket\psi = \ket{\psi'} - \ket\psi = - {1\over2}
  \left( \Ldag\L - \expect{\Ldag\L} \right) \ket\psi \delta t \;.
\label{nonlinear_schrod}
\end{equation}
Since the result 0 is highly probable, as the system interacts with
successive environment bits in initial state $\ket0_E$,
most of the time the system
state will evolve according to the nonlinear (and nonunitary) continuous
equation (\ref{nonlinear_schrod}).

Occasionally, however, a measurement result of 1 will be obtained.
In this case, the state of the system can change dramatically:
\begin{equation}
\ket\psi \rightarrow \ket{\psi'} =
  { \L\ket\psi \over \sqrt{\expect{\Ldag\L}}} \;.
\label{theJump}
\end{equation}
Because these changes are large but rare, they are usually referred to
as {\it quantum jumps}.

These two different evolutions---continuous and deterministic (after
a measurement result 0) or discontinuous and random (after a
measurement result 1)---can be combined into a single {\it stochastic
Schr\"odinger equation}:
\begin{eqnarray}
\delta\ket\psi = \ket{\psi'} - \ket\psi &=&
  - {1\over2}\left(\Ldag\L - \expect{\Ldag\L}\right) \ket\psi \delta t
  \nonumber\\
&& + \left( {\L\over \sqrt{\expect{\Ldag\L}}}
  - \id \right)\ket\psi \delta N \;,
\label{jump_eqn}
\end{eqnarray}
where $\delta N$ is a {\it stochastic variable} which is usually 0,
but has a probability $p_1 = \bra\psi\Ldag\L\ket\psi\delta t$
of being 1 during a given timestep $\delta t$.
The values of $\delta N$ obviously represent
measurement outcomes; when the environment is measured to be in state
$\ket0$ the variable is $\delta N=0$; when the measurement outcome is 1,
$\delta N=1$.  The equation (\ref{jump_eqn}) combines the two kinds of
system evolution into a single equation.  Most of the time the $\delta N$
term vanishes, and the system state evolves according to the
deterministic nonlinear equation (\ref{nonlinear_schrod}); however, when
$\delta N=1$ the $\delta N$ terms completely dominate, and the system state
changes abruptly according to (\ref{theJump}).  Generically, (\ref{jump_eqn})
can also include a term for Hamiltonian evolution, but this is
eliminated by the assumption (\ref{first_order}).

We can summarize the behavior of the stochastic variable by giving
equations for its ensemble mean,
\begin{equation}
(\delta N)^2 = \delta N \;, \ \ M[\delta N] = \expect{\Ldag\L}\delta t \;,
\end{equation}
where $M[]$ represents the ensemble mean over all possible measurement
outcomes, and $\expect{}$ is the quantum expectation value in the state
$\ket\psi$.

This analysis is rather abstract.  Let us see how this works with
a particular choice of two-bit interaction.  Suppose that the
system interacts with a succession of environment q-bits,
via the transformation $\Z_S(\theta)\U_{\rm CNOT}(\theta)$,
where $\Z_S(\theta) = \exp(-i\theta\sz/2)_S\otimes\id_E$
and $\theta \ll 1$; and that after each interaction the environment bits
are measured in the $z$ basis.  If the environment bits all begin in state
$\ket0$, and the system in state (\ref{qbit_pure}), after interacting with
the first environment q-bit the two bits will be in the joint state
\begin{equation}
\ket{\Psi'} = \alpha\ket{00} + \beta\cos\theta\ket{10}
  - i \beta\sin\theta\ket{11} \;,
\label{weakCNOT}
\end{equation}
modulo an overall phase.
This state is entangled, with entropy of entanglement
$S_E \approx - |\alpha\beta|^2\theta^2 \log_2|(\alpha\beta|^2\theta^2)$.
If the environment is then measured in the $z$ basis, it will be found in
state $\ket0$ with probability
$p_0 = |\alpha|^2 + |\beta|^2\cos^2\theta \approx 1 - |\beta|^2\theta^2$
and in state $\ket1$ only with the very small probability
$p_1 = |\beta|^2\theta^2$.  After a result of 0, the system will then be
in the new state
\begin{eqnarray}
\ket{\psi_0} &=& (\alpha\ket0
  + \beta\cos\theta\ket1)/\sqrt{p_0} \nonumber\\
&\approx& \alpha (1+|\beta|^2\theta^2/2) \ket0
  + \beta (1-|\alpha|^2\theta^2/2) \ket1 \;,
\end{eqnarray}
where we have kept terms up to second order in $\theta$.
The amplitude of $\ket0$ increases relative to $\ket1$.  On the other hand,
if the outcome 1 occurs, the system will be left in state $\ket1$.
Further interactions and measurements will not alter this.  We see that
this combination of interactions and measurements is exactly equivalent
to a weak measurement on the system, of the first type discussed in
section 3.2.
This bears out the fact that all POVMs are equivalent
to an interaction with an ancillary system followed by a projective
measurement, the well-known ``Neumark's Theorem.''

Suppose that the system interacts successively with $n$ environment
q-bits, each of which is afterwards measured to be in state $\ket0$.
The state of the system is $\ket{\psi_n} = \alpha_n\ket0 + \beta_n\ket1$,
where
\begin{equation}
{\beta_n\over\alpha_n} =
  \left( 1 - {\theta^2\over2} \right)^n
  {\beta\over\alpha} \approx
  \exp( - n\theta^2/2) {\beta\over\alpha} \;,
\end{equation}
which, together with the normalization condition
$|\alpha_n|^2+|\beta_n|^2=1$, implies
\begin{eqnarray}
|\alpha_n|^2
  &=& {|\alpha|^2\over |\alpha|^2
 + |\beta|^2\exp(-n\theta^2) } \;, \nonumber\\
|\beta_n|^2
  &=& {|\beta|^2\exp(-n\theta^2)\over |\alpha|^2
 + |\beta|^2\exp(-n\theta^2) } \;.
\label{asymptotic_jump}
\end{eqnarray}
We see that $|\alpha_n|^2 \rightarrow 1$, which is what one would expect,
since after many measurement outcomes without a single 1 result, one would
estimate that the $\ket1$ component of the state must be very small.
The probability of measuring 0 at the $n$th step, conditioned on having
seen 0s at all previous steps,
is $p_0(n) \approx 1 - |\beta_n|^2\theta^2 \rightarrow 1$.

\medskip\noindent
{\bf Exercise 6.} Show that the the probability of getting the result 0 at
every step as $n\rightarrow\infty$, which leads to the system asymptotically
evolving into the state $\ket0$, is
\begin{equation}
\prod_{n=0}^\infty p_0(n) \approx |\alpha|^2 \;.
\end{equation}
\medskip

The probability of at some point getting a result 1 and leaving the
system in state $\ket1$ is therefore $1-|\alpha|^2 = |\beta|^2$.
The effect, after many weak measurements, is exactly the same as a single
strong measurement in the $z$ basis, with exactly the same probabilities.

This evolution is described exactly by the equation (\ref{jump_eqn})
with Lindblad operator
\begin{equation}
\L = \sqrt{\theta^2/\delta t} \ket1\bra1 \;.
\label{CNOTlindblad}
\end{equation}
In this case, the equation simplifies to
\begin{equation}
\ket{\psi'}-\ket\psi = -{\theta^2\over2}
  \left( \ket1\bra1 - |\bracket{1}{\psi}|^2 \right) \ket\psi
  + \left( {\ket1\bra1\over|\bracket{1}{\psi}|}
  - \id \right)\ket\psi \delta N \;,
\end{equation}
where $M[\delta N] = \theta^2 |\bracket{1}{\psi}|^2 = \theta^2|\beta|^2 = p_1$.
Equation (\ref{jump_eqn}) may seem unnecessarily
complicated.  After all, $\L$ is Hermitian, so the distinction between
$\L$ and $\Ldag$ is unimportant; also, $\L$ is proportional to the
projector $\ket1\bra1$, which simplifies (\ref{jump_eqn}) considerably.
However, many other systems exist in which $\L$ is not so simple, and
which still obey equation (\ref{jump_eqn}).

We can readily find such an alternative system.
Suppose that $\U_{\rm SWAP}(\theta)$
is used instead of
$\U_{\rm CNOT}(\theta)$
in the two-bit interaction.  The system and environment bits
after the interaction are in the state
\begin{equation}
\ket{\Psi'} = \alpha\ket{00} + \beta\cos\theta\ket{10}
  - i \beta\sin\theta\ket{01} \;.
\label{weakSWAP}
\end{equation}
(Note that unlike the {\it strong} swap gate $\U_{\rm SWAP}$, the weak
$\U_{\rm SWAP}(\theta)$ {\it can} produce entanglement.)
From (\ref{weakSWAP}) we see that just as in the CNOT case,
a measurement of 0 on the environment leaves the system in a slightly
altered state---in fact, exactly the same state that is produced by a 0
result in the CNOT case.  The probabilities $p_0$ and $p_1$ for the two
outcomes are exactly identical to those for the CNOT case,
$p_0 = 1 - p_1 \approx 1 - |\beta|^2\theta^2$.  However, unlike
the CNOT case, a measurement outcome 1 will put the system
into the state $\ket0$ rather than $\ket1$.

This process is quite like spontaneous decay.  Suppose that $\ket0$
is the ground state and $\ket1$ is the excited state.  Initially, the
system is in a superposition of these two energy levels.  Each timestep,
there is a chance for the system to emit a quantum of energy into the
environment.  If it does, the system drops immediately into
the ground state $\ket0$, while the environment goes into the
excited state $\ket1$; if not, we revise our estimate of the system state,
making it more probable that the system is unexcited.  This is exactly
similar to quantum jumps in quantum optics, in which
a photodetector outside a high-$Q$ cavity clicks if a photon escapes,
and gives no output if it sees no photon \cite{Carmichael,Dum,Gardiner}.
A measurement result of 1 on the environment corresponds to the
photodetector click.

This model obeys exactly the same stochastic Schr\"odinger equation
(\ref{jump_eqn}), but with the Lindblad operator
$\L = \sqrt{\theta^2/\delta t}\ket0\bra1$.  Because this operator
is {\it not} Hermitian, the distinction between $\Ldag$ and $\L$ is important
in this case.
We compare these two stochastic evolutions in Fig.~3a and 3b.  In both
cases, the coefficient $|\beta|^2$ decays steadily away; however, in
some trajectories the state suddenly jumps either to $\ket1$ or $\ket0$,
while others continue to decay smoothly towards $\ket0$.

What happens if we don't actually measure the environment bits?  In this
case the system q-bit evolves into a mixed state which is the average over
all possible measurement outcomes.  If we denote by $\xi$ the $n$ outcomes
after $n$ steps of the evolution, $p(\xi)$ the probability of those outcomes,
and $\ket{\psi_\xi}$ the state conditioned on those outcomes (i.e., the
solution of the appropriate stochastic Schr\"odinger equation), then the
system state will be a density operator
\begin{equation}
\rho^{(n)} = \sum_\xi \ket{\psi_\xi} p(\xi) \bra{\psi_\xi} \;.
\end{equation}
If we start with a state $\rho$, allow it to interact with exactly
one environment q-bit, and average over the possible outcomes
0 and 1 with their correct probabilities, we will find that the new state
$\rho'$ is
\begin{equation}
\rho' = \sum_{k=0}^1 \A_k \rho \Adag_k \;,
\label{rhoprime}
\end{equation}
where for $\U_{\rm CNOT}$
\begin{eqnarray}
\A_0 &=& \ket0\bra0 + \cos\theta \ket1\bra1 \;, \nonumber\\
\A_1 &=& \sin\theta \ket1\bra1 \;.
\label{ucnot1}
\end{eqnarray}
In the limit of small $\theta\ll1$ this evolution gives an approximate
Lindblad master equation
\begin{equation}
{{\rho' - \rho}\over\delta t} =
  \L \rho \Ldag - {1\over2} \Ldag\L \rho
  - {1\over2} \rho \Ldag\L \;.
\label{mean_lindblad}
\end{equation}
The Lindblad operator $\L$ is the same as in the jump equation for
the CNOT interaction, (\ref{CNOTlindblad}).  Similarly, for
the SWAP interaction, the evolution equations are
identical, but with $\A_1 = \sin\theta \ket0\bra1$ and
$\L = \sqrt{\theta^2/\delta t}\ket0\bra1$.

We should note that unlike the stochastic trajectory evolution, the
evolution of the density matrix is perfectly deterministic, with no sign
of jumps or other discontinuities.  This is generally true; the jumps
appear only if measurements are performed.

Believe it or not, the models analyzed in this section have
already given almost all essential properties of quantum trajectories.
Quantum trajectory equations arise when a system interacts with an
environment which is subsequently measured; the equations give the
evolution of the system state conditioned on the measurement outcome.
They take the form of stochastic nonlinear differential equations; the
stochasticity arises due to the randomness of the measurement outcomes,
and the nonlinearity due to renormalization of the state.  Averaging
over all possible measurement outcomes recovers the deterministic
evolution of the system density operator.  The next two sections examine
elaborations of this basic scheme, and section 10 discusses how
these results fit into the foundations of quantum mechanics.

\section{Stochastic Schr\"odinger equations with \\ diffusion}

In discussing weak measurements we noted that some usually produced only
minor effects, but occasionally caused a large change in the state, while
others changed the state little regardless of the outcome.  The former
case is exactly that of trajectories with jumps:  the state usually
changes slowly and continuously, but occasionally makes a large jump.
The latter case, then, should correspond to states which evolve by small
but unpredictable changes---in other words, by diffusion rather than jumps.

\subsection{Random unitary diffusion}

We can easily demonstrate such evolution using the model we've already
constructed.  Suppose that the system begins in a state
(\ref{qbit_pure}), and interacts with a succession of environment bits
initially in state $\ket0$ by the unitary transformation
$\Z_S(\theta)\U_{\rm CNOT}(\theta)$ for $\theta\ll1$.
However, rather than measuring the environment bits in the $z$ basis,
we measure them in the $x$ basis,
\begin{equation}
\ket{x_+} = (\ket0 + \ket1)/\sqrt2 \;,\ \ 
\ket{x_-} = (\ket0 - \ket1)/\sqrt2 \;.
\end{equation}
In terms of this basis, after interaction the system and environment
bit are in the state
\begin{equation}
\ket{\Psi'} = {1\over\sqrt2}\left( \alpha\ket0 +
  \beta\e^{-i\theta}\ket1 \right) \ket{x_+} +
  {1\over\sqrt2}\left( \alpha\ket0 +
  \beta\e^{+i\theta}\ket1 \right) \ket{x_-} \;.
\end{equation}
We see that the two measurement outcomes $+$ and $-$ are equally likely,
and produce only a small effect on the system state:  namely, if an outcome
$+$ occurs the relative phase between $\ket1$ and $\ket0$ is rotated
by $-\theta$, and if a $-$ occurs it is rotated by $+\theta$.  This
measurement scheme, then, has the effect
of performing a random unitary transformation on the system.

The stochastic Schr\"odinger equation for this model is
\begin{equation}
\delta\ket\psi = \ket{\psi'} - \ket\psi = i \L \ket\psi \delta W \;,
\label{unitary_diffusion}
\end{equation}
where $\L=\sqrt{\theta^2/\delta t}\ket1\bra1$ is the same as in the
earlier case of CNOT jumps, and $\delta W$ is a stochastic variable taking
the values $\delta W = \pm \sqrt{\delta t}$ with equal probability.
We express this stochastic behavior by giving the ensemble means
\begin{equation}
M[\delta W] = 0 \;, \ \ \delta W^2 = \delta t \;.
\label{ensemble_means1}
\end{equation}
This particular normalization of the stochastic variable is useful
because it scales properly with time.  If, instead of using interactions
with single environment q-bits in intervals $\delta t$, we let
the system interact with $n$ bits over an interval $\Delta t = n\delta t$,
then the equation (\ref{unitary_diffusion}) remains completely
unchanged, except for $\delta t$ being replaced by $\Delta t$ in
(\ref{unitary_diffusion}) and (\ref{ensemble_means1}).

What if we average over {\it these} measurement outcomes?  In this case,
the density matrix evolution is still given by (\ref{rhoprime}),
with operators
\begin{eqnarray}
\A_0 &=& {1\over\sqrt2}\left(\ket0\bra0
  + \e^{-i\theta}\ket1\bra1 \right) \;, \nonumber\\
\A_1 &=& {1\over\sqrt2}\left(\ket0\bra0
  + \e^{+i\theta}\ket1\bra1 \right) \;.
\label{ucnot2}
\end{eqnarray}
This looks quite different from the evolution given by (\ref{ucnot1}),
but it is actually identical, as it must be:  the same system state must
be obtained by tracing out the environment bit, regardless of the basis
in which we choose to carry out the trace.
At the level of density matrices, the nonunitary quantum jumps and the
unitary diffusion are exactly the same.

\medskip\noindent
{\bf Exercise 7.}  Show explicitly that the new density matrix
\begin{equation}
\rho' = \sum_k \A_k \rho \Adag_k
\end{equation}
obtained using the operators defined in (\ref{ucnot2}) is the same as
that using the operators defined in (\ref{ucnot1}).
\medskip

This is an example of a single master equation having different
{\it unravelings}, that is, different stochastic evolutions
(\ref{jump_eqn}) and (\ref{unitary_diffusion}) which give
the same evolution on average.  Any set of measurements done on the
environment must give the same {\it average} evolution of the system state,
and therefore gives an unraveling of the same master equation.  Of
course, some unravelings may give simpler versions of the stochastic
system state evolution than others.  For some systems, it may be much
easier to solve the stochastic evolution than the master equation; one
can then average over many different stochastic evolutions to approximate
the master equation evolution.  This is called the Monte Carlo Wavefunction
technique \cite{Dalibard}.  Even in cases where one has no direct
access to information about the environment,
it may be worthwhile to invent a fictitious set
of measurements which would give a simple stochastic evolution, purely
as a numerical technique to solve the master equation.  In this case,
however, one should not ascribe any fundamental significance to the
individual solutions of the stochastic trajectory equations.

\subsection{Nonunitary diffusion}

We can get diffusive behavior from this system in a different way.
Suppose once again that the interaction is
$\Z_S(\theta)\U_{\rm CNOT}(\theta)$
and the environment is measured in the $z$ basis, but this time let the
environment q-bits be initially in the state
\begin{equation}
\ket{y_-} = (\ket0 - i\ket1)/\sqrt2 \;.
\end{equation}
In this case, the joint state after the interaction is
\begin{eqnarray}
\ket{\Psi'} &=& {1\over\sqrt2}(\alpha\ket0
  + \beta(\cos\theta-\sin\theta)\ket1 ) \ket0 \nonumber\\
&& - {i\over\sqrt2}(\alpha\ket0
  + \beta(\cos\theta+\sin\theta)\ket1 ) \ket1 \;.
\end{eqnarray}
For small $\theta\ll1$ the two measurement outcomes are almost equally
likely, $p_{0,1} = (1/2) \mp \theta|\beta|^2$, and the conditioned states
of the system are, to second order in $\theta$,
\begin{eqnarray}
\ket{\psi_0} &\approx&
 \alpha(1+|\beta|^2\theta+3|\beta|^4\theta^2/2)\ket0 \nonumber\\
&& + \beta(1-|\alpha|^2\theta
 - |\beta|^2\theta^2 - \theta^2/2 + 3|\beta|^4\theta^2/2)\ket1 \;, \nonumber\\
\ket{\psi_1} &\approx&
 \alpha(1-|\beta|^2\theta+3|\beta|^4\theta^2/2)\ket0 \nonumber\\
&& + \beta(1+|\alpha|^2\theta
 - |\beta|^2\theta^2 - \theta^2/2 + 3|\beta|^4\theta^2/2)\ket1 \;.
\end{eqnarray}
This is essentially the same as
the second weak measurement scheme (\ref{weak2}) in section 3.2.
If the system is initially in either state $\ket0$ or $\ket1$ it is unchanged,
but all other states will diffuse along the Bloch sphere until they eventually
reach either $\ket0$ or $\ket1$.  Once the state is within a small
neighborhood of $\ket0$ or $\ket1$ it is unlikely to diffuse away again.
Because of this, this scheme can also be thought of as an indirect
measurement of $\ket0$ vs. $\ket1$; but in this case, the measurement
result can only be determined by gathering a very large number of outcomes.
If, after the system has interacted with many environment bits, we
find that the number of 1 outcomes slightly exceeds the number of 0 outcomes,
we would conclude that the system is in state $\ket1$; if not, that the
system is in state $\ket0$.

This scheme is somewhat analogous to homodyne measurement in quantum optics
\cite{Carmichael,Wiseman}.  In homodyne detection, the weak signal from the
system is coherently mixed with a strong local oscillator in a beam splitter;
the two ports are fed into separate photodetectors.  The output from each
detector is almost entirely due to the local oscillator field; because
this field is so much stronger than the signal from the system, almost
all of the photons which arrive at the photodetectors come from the local
oscillator, and very little information about the system is obtained
in any one detector ``click.''  To gain information about the system, one
looks at the small difference between the photocurrents from the two detectors
over the course of many ``clicks.''  Similarly in our model,
over the course of many events one can deduce information
about the system state from the small excess of $1$ results over $0$ results.
The analogy to homodyne measurement is not as exact as the analogy between
the quantum jump equation and spontaneous emission, but it is still
suggestive.

For this nonunitary diffusion, the stochastic Schr\"odinger equation
is different than in the earlier quantum jump equation (\ref{jump_eqn}).
Expanding the solution to second order in $\theta$, the equation is
\begin{eqnarray}
\ket{\psi'} - \ket\psi &=&
  {1\over2}\left(3|\expect{\L}|^2 - 2\expect{\Ldag}\L
  - \Ldag\L\right)\ket\psi \delta t \nonumber\\
&& + (\L - \expect{\L})\ket\psi \delta W \;,
\end{eqnarray}
where once again $\L = \sqrt{\theta^2/\delta t}\ket1\bra1$, and
just as in the random unitary case of section 8.1 the stochastic
variable $\delta W$ takes the values $\mp \sqrt{\delta t}$ with 
probabilities $p_{0,1}$.
Because in this case the measurement outcomes are not equally probable,
this stochastic variable $\delta W$ does {\it not} have vanishing mean:
\begin{equation}
M[\delta W] = 2\expect{\Ldag}\delta t \;, \ \ \delta W^2 = \delta t \;.
\end{equation}
We can simplify this equation by replacing this stochastic variable by
one that does have a vanishing mean (to second order in $\theta$).  Define
\begin{eqnarray}
\delta Z &=& \delta W - M[\delta W] \;, \nonumber\\
M[\delta Z] &=& 0 \;, \ \ \delta Z^2 = \delta t + O(\delta t^{3/2}) \;.
\end{eqnarray}
Using this new stochastic variable the equation becomes
\begin{eqnarray}
\delta\ket\psi = \ket{\psi'} - \ket\psi &\approx& 
  \left( \expect{\Ldag}\L - {1\over2}|\expect{\L}|^2
  - {1\over2}\Ldag\L\right)\ket\psi \delta t \nonumber\\
&& + (\L - \expect{\L})\ket\psi \delta Z \;.
\label{real_qsd}
\end{eqnarray}
This is the {\it quantum state diffusion equation with real noise}
\cite{Gisin1}.  Solutions of this equation are plotted in Figure 4.

If we average over the measurement outcomes to give a mixed-state
evolution for the system, the new density matrix is given by
(\ref{rhoprime}) with operators
\begin{eqnarray}
\A_0 &=& {1\over\sqrt2} \left( \ket0\bra0
  + (\cos\theta-\sin\theta) \ket1\bra1 \right) \;, \nonumber\\
\A_1 &=& {1\over\sqrt2} \left(  \ket0\bra0
  + (\cos\theta+\sin\theta) \ket1\bra1 \right) \;.
\label{ucnot3}
\end{eqnarray}

\medskip\noindent
{\bf Exercise 8.}  Using the operators (\ref{ucnot3}), show that the
density operator evolution is exactly the same as that given by operators
(\ref{ucnot1}) and (\ref{ucnot2}), despite the fact that the environment's
initial state is quite different in this case, and that the quantum
trajectory evolution is also very different.  We can see from this that
the system evolution by itself is not sufficient to uniquely determine the
nature of the environment.  A particular master equation can result from
many different environments and interactions.
\medskip

\section{Incomplete information and stochastic \\ master equations}

So far, we have only considered models in which both the system and
environment are in a pure state, and measurements
are done which leave that the case.
The experimenter has complete information
about the state of the system and environment at all times.
It is possible to generalize this in a numbers of ways.  We may have only
partial information about the state of the system, so that it is in an
initial density matrix rather than a pure state.  We may have only partial
information about the environment; for instance, it might be in an initial
thermal state.  Or the measurements we perform on the environment may,
themselves, give only partial information.  In a realistic experiment,
of course, all three of these may be true.  We will consider each of
them in turn.

\subsection{Trajectories for a mixed system state}

A pure state represents the maximum information that one can possess about
a system.  If a system is in a pure state, that implies that there is
some measurement which could be performed on the system which would have
a definite outcome.  If we have only partial information about a system,
we describe it by a mixed state.  If a system is in a mixed state, there
is {\it no} measurement which has a definite outcome.  Systems can be
mixed because we lack information about how they were prepared, or because
they are entangled with other systems, or both.  The von~Neumann entropy
is the most useful measure of how mixed a state is; it vanishes for a
pure state, and is maximized by the maximally mixed state.

For simplicity, let's suppose that we have no information at all about
the state of the system; it is in the maximally mixed state
$\rho = \id/2$, and has von~Neumann entropy $S=1$.
We let it interact with an environment q-bit in some initial state,
and measure the environment bit in some basis afterwards.

For these measurement schemes, the state of the system becomes
\begin{equation}
\rho \rightarrow \rho_i = \A_i \rho \Adag_i / p_i \;,
\label{trajops}
\end{equation}
with probability
\begin{equation}
p_i = \tr\{\rho\Adag_i\A_i\} \;.
\end{equation}
If the interaction is $\Z_S(\theta)\U_{\rm CNOT}(\theta)$, as in the
first quantum jump equation of section 7, the environment bits are
initially in state $\ket0$, and are afterwards measured in the $z$ basis,
the operators (\ref{trajops}) are given by (\ref{ucnot1}).  Just as
for the pure state, an outcome 1 leaves the system in the state $\ket1$,
while an outcome 0 leaves the state only slightly changed:
\begin{eqnarray}
\rho_0 &\approx& {1\over2}\left( (1+\theta^2/2) \ket0\bra0
  + (1-\theta^2/2) \ket1\bra1 \right) \;,\ \ 
  p_0 \approx 1 - \theta^2/2 \;, \nonumber\\
\rho_1 &=& \ket1\bra1 \;, \ \ p_1 \approx \theta^2/2 \;.
\end{eqnarray}

In either case, the state is no longer completely mixed.  Some information
has been acquired about the system.  We can assess how much information
this is by calculating the {\it average} von~Neumann entropy ${\bar S}$
of the state after the measurement:
\begin{equation}
\Delta S = 1 - {\bar S} = 1 + \sum_i p_i \tr\{ \rho_i \log_2 \rho_i \} \;,
\label{deltas}
\end{equation}
which in the case of our jump trajectories is
$\Delta S \approx \theta^2/(2\ln2)$.
How much information does the measurement generate?
This is given by the Shannon entropy (\ref{shannon})
of the measurement,
\begin{equation}
S_{\rm meas} \approx - (\theta^2/2) \log_2\theta^2/2 \;.
\end{equation}
(Note that in \cite{Soklakov}, Soklakov and Schack call ${\bar S}$
the {\it average conditional entropy} and denote it ${\bar H}$,
and call $S_{\rm meas}$ the {\it average preparation information}
and denote it ${\bar I}$.)

We see that $S_{\rm meas} \ge \Delta S$, which of course makes sense;
we can never obtain more information on average about the system than
the measurement produces.  If we look at the ratio of the two, though,
\begin{equation}
{ S_{\rm meas} \over \Delta S } \approx 1 - \ln \theta^2/2 \;,
\label{ratio}
\end{equation}
we see that as $\theta^2 \rightarrow 0$ the information produced by the
measurement is much, much higher than our gain in knowledge about the
system.  Thus, most of the information we receive is just random noise,
telling us not about the state of the system but about the measurement
process itself.

If, instead of the jump description (\ref{ucnot1}) given by measuring
the environment in the $z$ basis, we use the unitary diffusive description
given by (\ref{ucnot2}) which arises from measuring the environment in the
$x$ basis, the situation is very different.  Because $\A_0$ and
$\A_1$ are in this case proportional to unitary operators, the initial
maximally-mixed state $\rho = \id/2$ is left completely unchanged by
this trajectory.  Absolutely no information is gained about the system,
$\Delta S = 0$.  This lack of information is not special to the
maximally-mixed state.  Because the two measurement outcomes are equally
likely, $p_0 = p_1 = 1/2$, each measurement result represents
$S_{\rm meas}=1$ bit of information, but this tells one nothing about
the state of the system; it is all random noise.

Consider now the second diffusive case described in section 8.2,
in which the environment bits begin
in the state $\ket{y_-}$ and, after interacting with the system, are
measured in the $z$ basis.  In this case, with evolution operators given
by (\ref{ucnot3}), an initially maximally-mixed state is left in one of
the states
\begin{equation}
\rho_{0,1} \approx {1\over2}\left( (1\pm\theta)\ket0\bra0
  + (1\mp\theta)\ket1\bra1 \right) \;,
\label{less_mixed}
\end{equation}
with probabilities
\begin{equation}
p_{0,1} \approx {1\over2}(1\pm\theta) \;.
\end{equation}

\medskip\noindent
{\bf Exercise 9.}  Show that, once again, most of the information
in this measurement outcome is random noise, with the ratio
$\Delta S/S_{\rm meas} \rightarrow 0$ as $\theta\rightarrow0$.
\medskip

\subsection{Trajectories for a mixed environment}

Suppose that, rather than the system being initially mixed, the environment
bits are in a mixed state $\rho_E$?  For simplicity let's choose
the diagonal state
\begin{equation}
\rho_E = w_0 \ket0\bra0 + w_1 \ket1\bra1 \;,
\end{equation}
with $w_0+w_1=1$.  Let an initially pure system in state (\ref{qbit_pure})
interact with the environment bit via
$\Z_S(\theta)\U_{\rm CNOT}(\theta)$.  The joint
state of the system and environment after the interaction will be
\begin{eqnarray}
\rho_{\rm tot} &=& |\alpha|^2 w_0 \ket{00}\bra{00} +
 \alpha\beta^* w_0 \cos\theta \ket{00}\bra{10} -
 i \alpha\beta^* w_0 \sin\theta \ket{00}\bra{11} \nonumber\\
&& + |\alpha|^2 w_1 \ket{01}\bra{01} -
 i \alpha\beta^* w_1 \sin\theta \ket{01}\bra{10} +
 \alpha\beta^* w_1 \cos\theta \ket{01}\bra{11} \nonumber\\
&& + \alpha^*\beta w_0 \cos\theta \ket{10}\bra{00} +
 i \alpha^*\beta w_1 \sin\theta \ket{10}\bra{01} \nonumber\\
&& + |\beta|^2(w_0 \cos^2\theta + w_1\sin^2\theta) \ket{10}\bra{10} \nonumber\\
&& + i |\beta|^2(w_1 - w_0) \cos\theta\sin\theta \ket{10}\bra{11} +
 i \alpha^*\beta w_0 \sin\theta \ket{11}\bra{00} \nonumber\\
&& + \alpha^*\beta w_1 \cos\theta \ket{11}\bra{01}
 - i |\beta|^2(w_1 - w_0) \cos\theta\sin\theta \ket{11}\bra{10} \nonumber\\
&& + |\beta|^2(w_1 \cos^2\theta + w_0\sin^2\theta) \ket{11}\bra{11} \;.
\end{eqnarray}

For small $\theta\ll1$, if we measure the environment in the $z$ basis
the system will be left in one of the states
\begin{equation}
\rho_{0,1} \approx
  (1-\theta^2|\beta|^2 w_{1,0}/w_{0,1}) \ket{\psi'}\bra{\psi'}
  + (\theta^2|\beta|^2 w_{1,0}/w_{0,1}) \ket1\bra1 \;,
\end{equation}
with probabilities
\begin{equation}
p_{0,1} \approx w_{0,1} \pm \theta^2|\beta|^2(w_1 - w_0) \;,
\label{mixed_probs}
\end{equation}
where
\begin{equation}
\ket{\psi'} = \alpha(1 + \theta^2|\beta|^2/2) \ket0 + 
 \beta(1 - \theta^2|\alpha|^2/2) \ket1 \;.
\end{equation}
These states $\rho_{0,1}$ are both mixed; the uncertainty in the
state of the environment has led to uncertainty in the state of the
system.  The system evolution is perturbed by noise from the
environment.

This evolution cannot be described by a stochastic Schr\"odinger
equation; rather, one must use a stochastic {\it master} equation,
which gives the evolution of a density matrix conditioned on the
measurement outcomes.  The master equation takes the form
\begin{eqnarray}
\rho'_S - \rho_S &=&
  - {1\over2} \left\{ \Ldag\L - \expect{\Ldag\L},\rho_S \right\}
  \left( { w_0\over p_0 } (1-F)
  + {w_1\over p_1} F \right) \delta t \nonumber\\
&& + \left[ \L \rho_S \Ldag - \expect{\Ldag\L}\rho_S \right]
  \left( { w_1\over p_0 } (1-F) + { w_0\over p_1 } F
  \right) \delta t \;.
\label{stoch_master}
\end{eqnarray}
Here $\{\A,\B\} = \A\B + \B\A$ is the anticommutator, and
$\expect\O = \tr\{\O\rho\}$.
The stochastic variable $F$ takes the value 0 with probability $p_0$
and 1 with probability $p_1$.  We see then that this equation
(\ref{stoch_master}) reduces to (\ref{mean_lindblad}) in the mean,
and hence gives a {\it partial} unraveling of that equation in terms
of mixed states.  When $w_1 \rightarrow 0$ this master equation exhibits
jump-like behavior; when $w_1 \sim w_0$ it is diffusive.
If $w_0 = w_1 = 1/2$ then (\ref{stoch_master}) reduces to the mean
equation (\ref{mean_lindblad}), since the measurements yield no information
whatsoever about the system.  (Note that this needn't be true for all
possible interactions.  For instance, if we use $\U_{\rm SWAP}(\theta)$
instead of $\U_{\rm CNOT}(\theta)$, it {\it is} possible to obtain
information about the system even using a maximally mixed environment,
as we see below.)

If the system interacts with $n$ environment bits, each of which is
subsequently measured in the $z$ basis giving a measurement record $\xi$,
the system will be left in the state
\begin{equation}
\rho_\xi = p_\xi \ket{\psi_n}\bra{\psi_n} + (1-p_\xi)\ket1\bra1 \;,
\label{rho_xi}
\end{equation}
where $p_\xi$ depends on the exact sequence of measurement results, and
$\ket{\psi_n}$ is given by (\ref{asymptotic_jump}) and obeys
$\ket{\psi_n}\rightarrow\ket0$ as $n\rightarrow\infty$.  The state will
tend to evolve towards either $\ket0\bra0$ or $\ket1\bra1$ unless
$w_0=w_1=1/2$.  So we see that even when a state becomes mixed, under some
circumstances it can evolve towards a pure state again.

If instead of $\U_{\rm CNOT}(\theta)$ the interaction were
$\U_{\rm SWAP}(\theta)$, the noise would affect the system in an
different way.  In this case the system will be left in one of the
states
\begin{eqnarray}
\rho_0 &=& {1\over p_0}\left[ w_0 \ket{\psi_0}\bra{\psi_0}
  + w_1 |\alpha|^2\sin^2\theta \ket1\bra1 \right] \;, \nonumber\\
\rho_1 &=& {1\over p_1}\left[ w_1 \ket{\psi_1}\bra{\psi_1}
  + w_0 |\beta|^2\sin^2\theta \ket0\bra0 \right] \;,
\end{eqnarray}
where
\begin{eqnarray}
\ket{\psi_0} &=& \alpha\ket0
  + \beta\cos\theta\ket1 \;, \nonumber\\
\ket{\psi_1} &=& \alpha\cos\theta\ket0
  + \beta\e^{-2i\theta}\ket1 \;,
\end{eqnarray}
and
\begin{equation}
p_0 = w_0 \cos^2\theta + |\alpha|^2\sin^2\theta = 1 - p_1 \;.
\end{equation}
In this case, rather than evolving eventually towards one of the pure
states $\ket0,\ket1$, the system will evolve towards the state
$w_0 \ket0\bra0 + w_1\ket1\bra1$, i.e., it will become identical with
the mixed state of the environment bits.  This is quite similar to the
process of thermalization, by which a system interacting with a heat
bath evolves towards equilibrium.

\subsection{Trajectories with generalized measurements}

Now we will assume that both the system and environment q-bits begin
in a pure state, but that the measurement performed on the environment
bit is no longer a projective measurement, but instead a POVM which
provides only partial information about the state of the environment bit.

Let us consider a system where the system and environment q-bits interact
via $\Z_S(\theta)\U_{\rm CNOT}(\theta)$, the environment bits start initially
in the state $\ket0$, and the system is in state $\alpha\ket0+\beta\ket1$.
After the system and environment interact, the environment is measured
with a POVM using operators
\begin{eqnarray}
\E_0 \equiv q \ket0\bra0 + (1-q) \ket1\bra1 = \A_0^2 \;, \nonumber\\
\E_1 \equiv (1-q) \ket0\bra0 + q \ket1\bra1 = \A_1^2 \;, \nonumber\\
\A_0 \equiv \sqrt{q} \ket0\bra0 + \sqrt{1-q} \ket1\bra1 \;, \nonumber\\
\A_1 \equiv \sqrt{1-q} \ket0\bra0 + \sqrt{q} \ket1\bra1 \;.
\end{eqnarray}
Assume that $1/2 < q \le 1$.  A measurement result 0 then favors the
environment q-bit being in state $\ket0$, and a result 1 favors $\ket1$,
but neither result is conclusive.

After the system and environment interact, the probabilities of the
two outcomes are
\begin{equation}
p_0 = q (|\alpha|^2 + |\beta|^2 \cos^2\theta)
  + (1-q) |\beta|^2 \sin^2\theta = 1 - p_1 \;.
\end{equation}
If we assume $\theta\ll1$, tracing out the environment leaves the
system in one of the mixed states
\begin{eqnarray}
\rho_0 &\approx&
  \left( 1 - (1-q)|\beta|^2\theta^2/q \right) \ket{\psi'}\bra{\psi'}
  + \left( (1-q)|\beta|^2\theta^2/q \right) \ket1\bra1 \;, \nonumber\\
\rho_1 &\approx&
  \left( 1 - q|\beta|^2\theta^2/(1-q) \right) \ket{\psi'}\bra{\psi'}
  + \left( q|\beta|^2\theta^2/(1-q) \right) \ket1\bra1 \;,
\end{eqnarray}
with the updated state
\begin{equation}
\ket{\psi'} = \alpha(1+|\beta|^2\theta^2/2)\ket0
  + \beta(1-|\alpha|^2\theta^2/2)\ket1 \;.
\end{equation}
That is, the system state remains entangled with the environment q-bit.

This environmental interaction and measurement leaves the system in
a state of the form
\begin{equation}
\rho_S = w \ket{\psi'}\bra{\psi'} + (1-w) \ket1\bra1 \;.
\end{equation}
If a state in that form interacts with another environment bit, on which
the same POVM is performed, the system state will remain in the same form.
After $n$ interactions, the state will have the form (\ref{rho_xi}),
with $\ket{\psi_n}\rightarrow\ket0$ as $n\rightarrow\infty$, and the
state tending towards either $\ket0\bra0$ or $\ket1\bra1$ at long times.
Thus, the system evolution looks quite similar to the case when the
environment is initially in a mixed state and a projective measurement
is performed.  However, in the case of a mixed environment the system
has no entanglement with the environment, and is in a mixed state due to
noise; in the case of incomplete measurements, the system is entangled
with the environment, and the system and environment as a whole remain
in a pure state.

In this case as well, one can describe the evolution of the system by
a stochastic master equation.  This has exactly the same form as
(\ref{stoch_master}), with $w_{0,1}$ replaced by $q$ and $1-q$.  Unlike
the previous case, however, the system state is not mixed because of
classical noise entering from the environment, but rather because it
remains entangled with the environment.

The POVM just considered represents a measurement with imperfect
discrimination.  Let's consider a different way that a measurement may
be incomplete.  Consider the following positive operators:
\begin{eqnarray}
\E_0 &\equiv& q \ket0\bra0 = \A_0^2 \;, \nonumber\\
\E_1 &\equiv& q \ket1\bra1 = \A_1^2 \;, \nonumber\\
\E_2 &\equiv& (1-q) \id = \A_2^2 \;, \nonumber\\
\A_0 &\equiv& \sqrt{q} \ket0\bra0 \;, \nonumber\\
\A_1 &\equiv& \sqrt{q} \ket1\bra1 \;, \nonumber\\
\A_2 &\equiv& \sqrt{1-q} \id \;.
\end{eqnarray}
These operators $\E_i$ sum up to the identity, and for $0\le q\le1$
form a POVM.  $\E_0$ and $\E_1$ are
proportional to orthogonal projectors; $\E_2$, by contrast,
leaves the state completely unchanged.  This is like a measurement made
with a detector of efficiency $q$; there is a probability $1-q$ that
the detector will fail to register anything at all, and hence gives
no information about the state.

Suppose the system q-bit is originally in a state of form
\begin{equation}
\rho_S = w_0 \ket\psi\bra\psi + w_1 \ket1\bra1 \;.
\label{rho_sys}
\end{equation}
(where an initial pure state would have $w_0=1$).  Then a measurement result
of 0 has probability
\begin{equation}
p_0 = q \left[ w_0(|\alpha|^2+|\beta|^2\cos^2\theta)
  + w_1\cos^2\theta \right] \approx q[1 - (1-w_0|\alpha|^2)\theta^2] \;,
\end{equation}
and leaves the system in a mixed state of the same form
\begin{equation}
\rho_0 \approx w_0(1+w_1|\alpha|^2\theta^2)\ket{\psi'}\bra{\psi'}
  + w_1(1-w_0|\alpha|^2\theta^2) \ket1\bra1 \;;
\end{equation}
the result 1 has probability
\begin{equation}
p_1 = q \left[ w_0|\beta|^2\sin^2\theta + w_1\sin^2\theta \right]
  \approx q (1-w_0|\alpha|^2)\theta^2 \;,
\end{equation}
and leaves the system in the state $\rho_1 = \ket1\bra1$; and the result
(or rather, nonresult) 2 has probability $p_2 = 1-q$ and leaves the
system in the mixed state
\begin{equation}
\rho_2 \approx w_0(1-|\beta|^2\theta^2)\ket{\psi'}\bra{\psi'}
  + (w_1 + w_0|\beta|^2\theta^2)\ket1\bra1 \;.
\end{equation}
If the evolution began in a pure state, it will remain in a pure state
unless outcome 2 occurs; but once this happens, the state will tend to
remain mixed until either a result 1 occurs (in which case the system
will be in state $\ket1$ thenceforth) or, in the limit of large $n$,
the system approaches the state $\ket0$.  The measurement process works
much as before, but the null results slow the convergence; if the efficiency
$q$ is low, it can take far longer for the state to settle down to
either $\ket0$ or $\ket1$.

\section{Quantum trajectories, consistent \\ histories, and collapse models}

As is clear from the previous sections, the formalism of quantum
trajectories calls on nothing more than standard quantum mechanics,
and as framed above is in no way an alternative theory or interpretation.
Everything can be described solely in terms of measurements and unitary
transformations, the building blocks of the usual Copenhagen interpretation.

However, many people have expressed dissatisfaction with the
standard interpretation over the years, usually due to
the role of measurement as a fundamental
building block of the theory.  Measuring devices are large, complicated
things, very far from elementary objects; what exactly constitutes a
measurement is never defined; and the use of classical mechanics to
describe the states of measurement devices is not justified.  Presumably
the individual atoms, electrons, photons, etc., which make up a detector
can themselves be described by quantum mechanics.  If this is carried to
its logical conclusion, however, and a Schr\"odinger equation is
constructed for the measurement process, one obtains not classical
behavior, but rather giant macroscopic superpositions such as the famous
Schr\"odinger's cat paradox \cite{Schroedinger}.

Three major approaches have been followed in tackling this problem.  In the
first, ``hidden variables'' are introduced which determine the outcome of
measurements, so that quantum mechanics is a partial description of an
underlying deterministic theory.  Unfortunately, as John Bell showed
\cite{Bell}, such a hidden variables theory
(such as that of de Broglie and Bohm \cite{Bohm}) must
be nonlocal to give the same predictions as quantum mechanics.

The second approach is to modify quantum mechanics to get rid of the
unwanted macroscopic superpositions, while retaining the usual quantum
results on small scales.  Theories of this nature have been proposed by
Pearle \cite{Pearle}, by Ghirardi, Rimini and Weber \cite{GRW},
by Di\'osi \cite{Diosi}, by Gisin \cite{Gisin1} and Percival \cite{Percival2},
and by Penrose \cite{Penrose}, among others; these theories are commonly
called ``collapse models.''

These alternative theories usually replace the Schr\"odinger equation
with a new, stochastic equation, which includes a mechanism to produce
wave function collapse onto some preferred basis.  Such stochastic
Schr\"odinger equations often strongly resemble those produced by a
suitable quantum trajectory description, in which the system interacts
continuously with an environment which is repeatedly measured.  While
I do not know if {\it all} such alternative theories are equivalent to a
quantum trajectory description, certainly a large class of them is.

Because these theories are not equivalent to standard quantum mechanics,
in principle they can be distinguished experimentally.  Unfortunately,
most such theories require experimental sensitivities well beyond current
technology, though Percival \cite{Percival2,Percival3}
has made interesting proposals of possible near-term tests.

Theoretically there are also problems in producing relativistic versions
of such theories.  Most such models produce collapse onto localized
variables, such as position states.  Such localization, however, does not
in general commute with the Hamiltonian, and so such theories almost
always violate energy conservation---sometimes spectacularly so
\cite{Adler}.  Another class of models produces a collapse
onto states of definite energy \cite{Diosi,Penrose,Percival2},
rather than localized variables.  It is not clear, however, that
such collapse models eliminate macroscopic superpositions in either the
short or long term.  Certainly the end result, a universe in an energy
eigenstate---a static, unchanging state with no dynamics---does not
greatly resemble the world that we perceive, though in the context of an
expanding universe even this is unclear.

The last approach is to retain the usual quantum theory, but to eliminate
measurement as a fundamental concept, finding some other interpretation
for the predicted probabilities.  While there are many interpretations
that follow this approach, the one that is most closely tied to quantum
trajectories is the {\it decoherent} (or {\it consistent}) {\it histories
formalism} of Griffiths, Omn\'es, Gell-Mann and Hartle
\cite{Griffiths,Omnes,GellMann}.  In this formalism,
probabilities are assigned to {\it histories} of events rather than
measurement outcomes at a single time.  These can be
grouped into {\it sets} of mutually exclusive, exhaustive histories whose
probabilities sum to 1.  However, not all histories can be assigned
probabilities under this interpretation; only histories which lie in sets
which are {\it consistent}, that is, whose histories do not
exhibit interference with each other, and hence obey the usual
classical probability sum rules.

Each set is basically a choice of description for the quantum system.
For the models considered in this paper, the quantum trajectories correspond
to histories in such a consistent set.  The probabilities of the
histories in the set exactly equal the probabilities of the measurement
outcomes corresponding to a given trajectory.  This equivalence has been
shown between quantum trajectories and consistent sets for certain more
realistic systems, as well \cite{DGHP,Brun1,Brun2}.

For a given quantum system, there can be multiple consistent descriptions
which are {\it incompatible} with each other; that is, unlike in classical
physics, these descriptions cannot be combined into a single, more
finely-grained description.  In quantum trajectories, different unravelings
of the same evolution correspond to such incompatible descriptions.
In both cases, this is an example of the complementarity of quantum
mechanics.

We see, then, that while quantum trajectories can be straightforwardly
defined in terms of standard quantum theory when the environment is
subjected to repeated measurements, even in the absence of such measurements
there is an interpretation of the trajectories in terms of decoherent
histories.  Because the consistency conditions guarantee that the
probability sum rules are obeyed, one can therefore use quantum trajectories
as a calculational tool even in cases where no actual measurements
take place.

\section{Conclusions}

In this paper I have presented a simple model of a system and environment
consisting solely of quantum bits, using no more than single-bit
measurements and one- and two-bit unitary transformations.  The
simplicity of this model makes it particularly suitable for demonstrating
the properties of quantum trajectories.  We see that these require no
more than the usual quantum formalism, though they can, of course, also
be applied in cases going beyond standard quantum mechanics, such as
collapse models or the decoherent histories interpretation.

Quantum trajectories can often simplify the description of an open
quantum system in terms of a stochastically evolving pure state rather
than a density matrix.  While for the q-bit models of this paper there
is no great advantage in doing so, for more complicated systems this
can often make a tremendous difference \cite{Schack}.

In this model it is also possible to straightforwardly quantify the
flow of information between system and environment, in the form of
entanglement and noise, as well as the information acquired in measurements
and the decrease in entropy of the system.  For realistic systems
and environments this may be more difficult to do analytically, but one
can still form a qualitative picture of the information flow in
decoherence and measurement.

I hope that this model will clear up much of the confusion that currently
surrounds the theory of quantum trajectories.  Quantum trajectories are
a very useful formal and numerical tool, and should find applications well
outside their current niches in quantum optics and measurement theory.

\section*{Acknowledgments}

The idea for this paper arose during a very productive visit to the University
of Oregon, and it has been encouraged by the kindness of many people.
I would particularly like to thank Steve Adler, Howard Carmichael,
Oliver Cohen, Lajos Di\'osi, Nicolas Gisin, Bob Griffiths, Jonathan Halliwell,
Jim Hartle, Jens Jensen, Ian Percival, Martin Plenio, R\"udiger Schack,
Artur Scherer, Andrei Soklakov, Tim Spiller and Dieter Zeh, for
their comments, feedback, and criticisms of the ideas behind this paper.
This work was supported in part by NSF Grant No.~PHY-9900755, by
DOE Grant No.~DE-FG02-90ER40542, and by the Martin A. and Helen Chooljian
Membership in Natural Sciences, IAS.

\section*{Appendix:  Solutions to Exercises}

{\bf Exercise 1.}  Any unitary operator can be written
$\U = \exp(i\H)$ where $\H$ is some Hermitian operator.  Any operator on
a single q-bit can be written in terms of the basis $\id,\sx,\sy,\sz$:
\[
\H = c_0 \id + c_1 \sx + c_2 \sy + c_3 \sz \;.
\]
For $\H=\H^\dagger$ to be Hermitian, the $c_i$ must all be real.  Choose
a convenient parametrization $c_{1,2,3} = \phi n_{x,y,z}$, where
$n_x^2 + n_y^2 + n_z^2 = 1$.  The expression for $\U$ is then
\[
\U = \exp(i c_0 \id + i\phi\nvec\cdot\svec)
  = \exp(i c_0 \id) \exp(i\phi\nvec\cdot\svec) \;,
\]
where the $c_0$ term factors out because $\id$ commutes with everything
else.  This factor just contributes a global phase change, and hence
can be neglected.  The second exponential can be expanded in a power
series to yield
\[
\exp(i\phi\nvec\cdot\svec)
  = \sum_{m=0}^\infty (i\phi\nvec\cdot\svec)^m/m! \;.
\]
Note now that
\begin{eqnarray}
(\nvec\cdot\svec)^2 &=& (n_x^2+n_y^2+n_z^2) \id
  + n_x n_y (\sx\sy + \sy\sx) \nonumber\\
&& + n_y n_z (\sy\sz + \sz\sy)
  + n_z n_x (\sz\sx + \sx\sz) \nonumber\\
&=& \id \, \nonumber
\end{eqnarray}
where we've used the facts that $\sx,\sy,\sz$ anticommute and that
$\sx^2 = \sy^2 = \sz^2 = \id$.  Substituting this into the power series
yields
\begin{eqnarray}
\U &=& \sum_{m=0}^\infty (i\phi)^{2m} (\nvec\cdot\svec)^{2m}/(2m)!
 + (\nvec\cdot\svec) \sum_{m=0}^\infty (i\phi)^{2m+1}
  (\nvec\cdot\svec)^{2m}/(2m+1)! \nonumber\\
&=& \id \sum_{m=0}^\infty (i\phi)^{2m}/(2m)!
 + (\nvec\cdot\svec) \sum_{m=0}^\infty (i\phi)^{2m+1}/(2m+1)! \nonumber\\
&=& \id \cos\phi + i(\nvec\cdot\svec)\sin\phi \;, \nonumber
\end{eqnarray}
which was to be shown.

{\bf Exercise 2.}  The proof of this is just the same as in Exercise 1.
Simply expand the exponential in powers of its argument
\[
\U(\theta) = \exp(-i\theta\U) = \sum_{m=0}^\infty (-i\theta\U)^m/m! \;.
\]
Because $\U^2=\Udag\U=\id$, all the odd
powers will be proportional to $\U$ and the even terms to $\id$.
Collecting the even and odd powers separately gives the result
\[
\U(\theta) = \U \cos\theta - i \U \sin\theta \;,
\]
just as in Exercise 1 above.

{\bf Exercise 3.}  The average information gain from a
measurement is given by the Shannon entropy (\ref{shannon}).
\[
S_{\rm meas} = - p_0 \log_2 p_0 - p_1 \log_2 p_1 \;.
\]
In this case, the relevant probabilities are $p_0 = 1-\epsilon|\beta|^2$
and $p_1 = \epsilon|\beta|^2$.  If we rewrite $\log_2 x = \ln x/\ln 2$,
we can use the expansion $\ln(1+\delta) = \delta - \delta^2/2
+ \delta^3/3 - \cdots$ to get
\[
S_{\rm meas}
  \approx  \epsilon|\beta|^2(1/\ln 2 - \log_2 \epsilon|\beta|^2) \;.
\]

{\bf Exercise 4.}  Because $\rho=\rho^\dagger$ is an Hermitian operator,
we can always find an orthonormal basis which diagonalizes it:
\[
\rho = \pmatrix{\lambda_1&0&0&\cdots&0\cr
  0&\lambda_2&0&\cdots&0\cr
  0&0&\lambda_3&\cdots&0\cr
  \vdots&\vdots&\vdots&\ddots&\vdots\cr
  0&0&0&\cdots&\lambda_n\cr} \;,
\]
where the $\{\lambda_i\}$ are all real.  The positivity of $\rho$ requires
that $\lambda_i \ge 0$, and $\tr\rho = 1$ implies $\sum\lambda_i = 1$.
Putting all this together, we find that the von~Neumann entropy becomes
\[
-\tr\{\rho\log_2\rho\} = - \sum_{i=1}^n \lambda_i \log_2 \lambda_i \;.
\]
The right-hand side of this equation is exactly the expression for a
Shannon entropy, as in (\ref{shannon}).  For an $n$-outcome measurement,
the Shannon entropy has a single global maximum at
$\lambda_1 = \lambda_2 = \cdots = \lambda_n = 1/n$.  If we substitute
these values into the matrix expression for $\rho$ above, we find that
$\rho = \id/n = \id/\tr{\id}$.

{\bf Exercise 5.}
\begin{eqnarray}
\sum_k \Odag_k\O_k &=&
  \sum_{jj'k} \lambda_k \mu_{kj}\mu^*_{kj'} \Adag_{j'}\A_j \nonumber\\
&=&  \sum_{jj'} M_{jj'} \Adag_{j'}\A_j \nonumber\\
&=&  \sum_{jj'} \Adag_{j'}\A_j \bra{E}\Bdag_{j'}\B_j\ket{E} \nonumber\\
&=&  \sum_{jj'} \tr_{\rm env} \left\{ (\Adag_{j'}\A_j \otimes \Bdag_{j'}\B_j)
  \left(\id_S\otimes\ket{E}\bra{E}\right) \right\}  \nonumber\\
&=&  \tr_{\rm env} \left\{ \left( \sum_{j'} \Adag_{j'}\otimes\Bdag_{j'}\right)
  \left(\sum_j \A_j\otimes\B_j\right) 
  \left( \id_S\otimes\ket{E}\bra{E} \right) \right\}  \nonumber\\
&=& \tr_{\rm env} \left\{ \Udag\U
  \left(\id_S\otimes\ket{E}\bra{E}\right) \right\}  \nonumber\\
&=& \tr_{\rm env} \left\{
  \id_S\otimes\ket{E}\bra{E} \right\}  \nonumber\\
&=& \id_S \;. \nonumber
\end{eqnarray}

{\bf Exercise 6.}  A very direct way of showing this result is
to use the simple approximation $1-\theta^2 \approx \exp(-\theta^2)$
for $\theta \ll 1$.  This gives:
\begin{eqnarray}
\prod_{n=0}^\infty p_0(n)
  &\approx& \prod_{n=0}^\infty (1-|\beta_n|^2\theta^2) \nonumber\\
&=& \prod_{n=0}^\infty {|\alpha|^2 + (1-\theta^2)|\beta|^2\exp(-n\theta^2)
  \over |\alpha|^2 + |\beta|^2\exp(-n\theta^2) } \nonumber\\
&\approx& \prod_{n=0}^\infty {|\alpha|^2 + |\beta|^2\exp(-(n+1)\theta^2)
  \over |\alpha|^2 + |\beta|^2\exp(-n\theta^2) } \nonumber\\
&=& \lim_{n\rightarrow\infty} {|\alpha|^2 + |\beta|^2\exp(-(n+1)\theta^2)
  \over |\alpha|^2 + |\beta|^2 } \nonumber\\
&=& |\alpha|^2 \;. \nonumber
\end{eqnarray}

A slightly more indirect approach is to take the natural logarithm
of the first line of the above equation, and use the approximation
$\ln(1+x) \approx x$ for $x \ll 1$.  This gives
\[
\ln \prod_{n=0}^\infty p_0(n)
  \approx \ln \prod_{n=0}^\infty (1-|\beta_n|^2\theta^2)
  \approx - \sum_{n=0}^\infty |\beta_n|^2\theta^2 \;.
\]
By approximating the sum as an integral and then exponentiating both
sides, one straightforwardly gets the same result.

{\bf Exercise 7.}  Using the definition (\ref{ucnot2}), we see that
\begin{eqnarray}
\sum_k \A_k \rho \Adag_k &=&
  {1\over2}\biggl( (\ket0\bra0 + \e^{-i\theta}\ket1\bra1 )
  \rho (\ket0\bra0 + \e^{+i\theta}\ket1\bra1 ) \nonumber\\
&& + (\ket0\bra0 + \e^{+i\theta}\ket1\bra1 )\rho
  (\ket0\bra0 + \e^{-i\theta}\ket1\bra1) \biggr) \;. \nonumber
\end{eqnarray}
By rearranging and combining terms, we get
\begin{eqnarray}
\sum_k \A_k \rho \Adag_k &=&
  {1\over2}\biggl( (\ket0\bra0 + (\cos\theta-i\sin\theta)\ket1\bra1 )
  \rho \nonumber\\
&& \times (\ket0\bra0 + (\cos\theta+i\sin\theta)\ket1\bra1 ) \nonumber\\
&& + (\ket0\bra0 + (\cos\theta+i\sin\theta)\ket1\bra1 )\rho \nonumber\\
&& \times (\ket0\bra0 + (\cos\theta-i\sin\theta)\ket1\bra1) \biggr) \nonumber\\
&=&   (\ket0\bra0 + \cos\theta\ket1\bra1)
  \rho (\ket0\bra0 + \cos\theta\ket1\bra1) \nonumber\\
&& + \sin^2\theta \ket1\bra1 \rho \ket1\bra1 \;, \nonumber
\end{eqnarray}
where the last equality is just $\A_0\rho\Adag_0 + \A_1\rho\Adag_1$ using
the definition (\ref{ucnot1}).

{\bf Exercise 8.}  The procedure here is just the same as in Exercise 7.
We write the new state in terms of the operators (\ref{ucnot3}) and
rearrange the terms to get the operators (\ref{ucnot1}).
\begin{eqnarray}
\rho' &=& \sum_k \A_k \rho \Adag_k \nonumber\\
&=& {1\over2}\biggl( (\ket0\bra0 + (\cos\theta-\sin\theta)\ket1\bra1 )
  \rho (\ket0\bra0 + (\cos\theta-\sin\theta)\ket1\bra1 ) \nonumber\\
&& + (\ket0\bra0 + (\cos\theta+\sin\theta)\ket1\bra1 )\rho
  (\ket0\bra0 + (\cos\theta+\sin\theta)\ket1\bra1) \biggr) \nonumber\\
&=& \ket0\bra0\rho\ket0\bra0 + 2\cos\theta(\ket0\bra0\rho\ket1\bra1
  + \ket1\bra1\rho\ket0\bra0 ) \nonumber\\
&& + \cos^2\theta \ket1\bra1\rho\ket1\bra1
  + \sin^2\theta \ket1\bra1\rho\ket1\bra1 \nonumber\\
&=& (\ket0\bra0 + \cos\theta\ket1\bra1) \rho
  (\ket0\bra0 + \cos\theta\ket1\bra1) \nonumber\\
&& + (\sin\theta\ket1\bra1)\rho(\sin\theta\ket1\bra1) \;. \nonumber
\end{eqnarray}
Of course, the evolutions given by (\ref{ucnot1}) and (\ref{ucnot2})
have already been shown to be equivalent, so the equivalence of
(\ref{ucnot3}) and (\ref{ucnot2}) follows at once.

{\bf Exercise 9.}  In this measurement, the system goes from the maximally
mixed state $\rho = \id/2$ to one of the two states $\rho_{0,1}$ given
by (\ref{less_mixed}).  The von~Neumann entropy of the maximally mixed
state is
\[
-\tr\left\{ {\id\over2}\log_2{\id\over2} \right\} =
  - {1\over2}\log_2{1\over2} - {1\over2}\log_2{1\over2} = 1 \;.
\]
Because the states $\rho_{0,1}$ are diagonal, it is straightforward to
evaluate their von~Neumann entropy as well; we find it to be
\[
-\tr\{\rho_{0,1}\log_2\rho_{0,1}\} = - p_0 \log_2 p_0 - p_1 \log_2 p_1
  = S_{\rm meas} \;,
\]
where $p_{0,1} = (1 \pm \theta)/2$.  So we see that
\[
S_{\rm meas} = 1 - {1\over2}\log_2(1+\theta) - {1\over2}\log_2(1-\theta)
  \approx 1 - \theta^2/(2\ln2) \;,
\]
which implies that $\Delta S \approx \theta^2/(2\ln2)$ and
$\Delta S/S_{\rm meas} \rightarrow 0$ as $\theta \rightarrow 0$, as required.
The entropy $S_{\rm meas}$ of the measurement is nearly a full bit, but
only a tiny part of it represents information about the system.

\eject
\vfil

\begin{figure}
\begin{center}
\epsfig{file=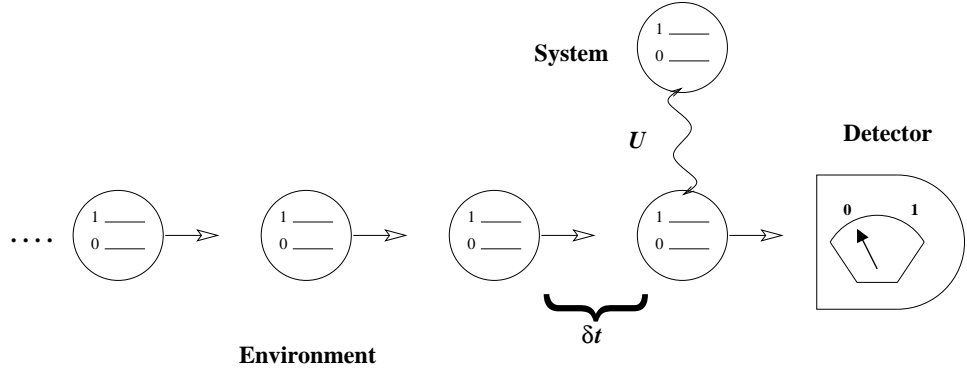, width=5in}
\label{fig1}
\end{center}
\caption{This is a schematic diagram of the type of model used in
this paper.  The system is a single two-level system, or q-bit.  It
interacts briefly with a series of environment q-bits which pass by
with average time intervals of $\delta t$, and the environment bits
are subsequently measured.}
\end{figure}

\eject
\vfil

\begin{figure}
\begin{center}
\epsfig{file=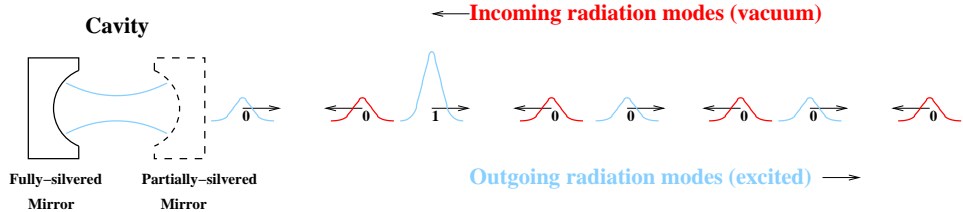, width=5in}
\label{fig2}
\end{center}
\caption{Here is a particular physical system of which the model
of this paper is an idealization.  The system is an electromagnetic
field mode in a cavity with a partially-silvered mirror.  There is on
average less than a single photon in the mode at a time, so only the
lowest two occupation levels contribute significantly to the dynamics.
The external electromagnetic field is described in terms of incoming
and outgoing wave packets of width $\delta t$, each of which reflects
on the outside of the partially-silvered mirror.  The incoming packets
are all in the vacuum state, but upon reflection they may absorb a
photon emitted from the cavity.}
\end{figure}

\eject
\vfil

\begin{figure}
\begin{center}
\epsfig{file=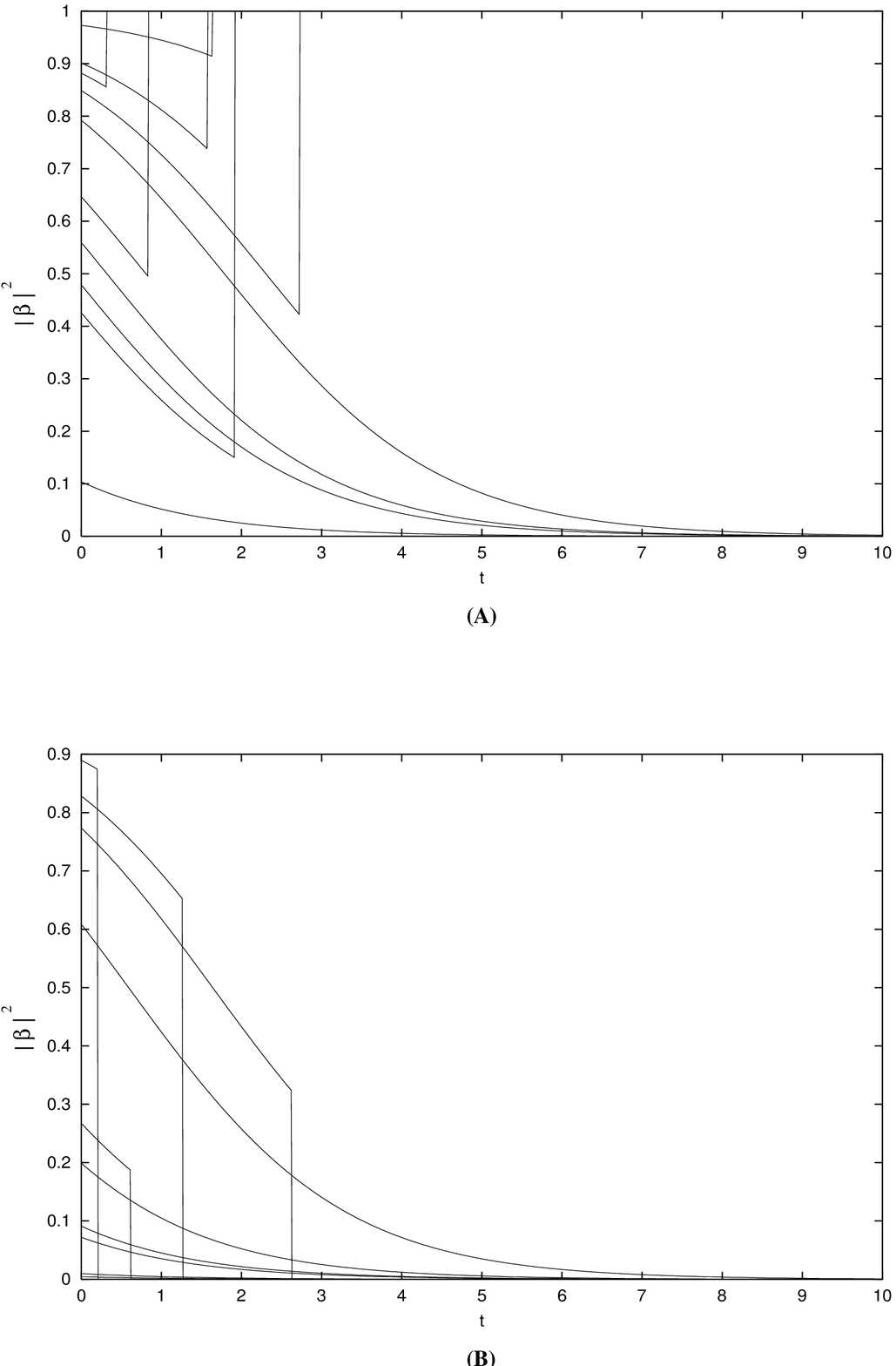, height=5in}
\label{fig3}
\end{center}
\caption{These two figures show a variety of quantum jump trajectories
for different initial states, plotting the squared amplitude $|\beta|^2$
for the system to be in state $\ket1$ versus the time.
Figure a) assumes an interaction
$\U_{\rm CNOT}$ with the environment, while b) assumes $\U_{\rm SWAP}$.
In both cases, $|\beta|^2$ decays towards zero.  Some trajectories just
decay smoothly to 0, while others undergo a jump.  In figure a) this
jump is to 1, while in b) they jump to 0, as in spontaneous emission.
Note that the decays of $|\beta|^2$ is {\it not} a simple exponential,
due to the nonlinearity of the trajectory equations.  As
$|\beta|^2 \rightarrow 0$, however, it approaches an exponential.}
\end{figure}

\eject
\vfil

\begin{figure}
\begin{center}
\epsfig{file=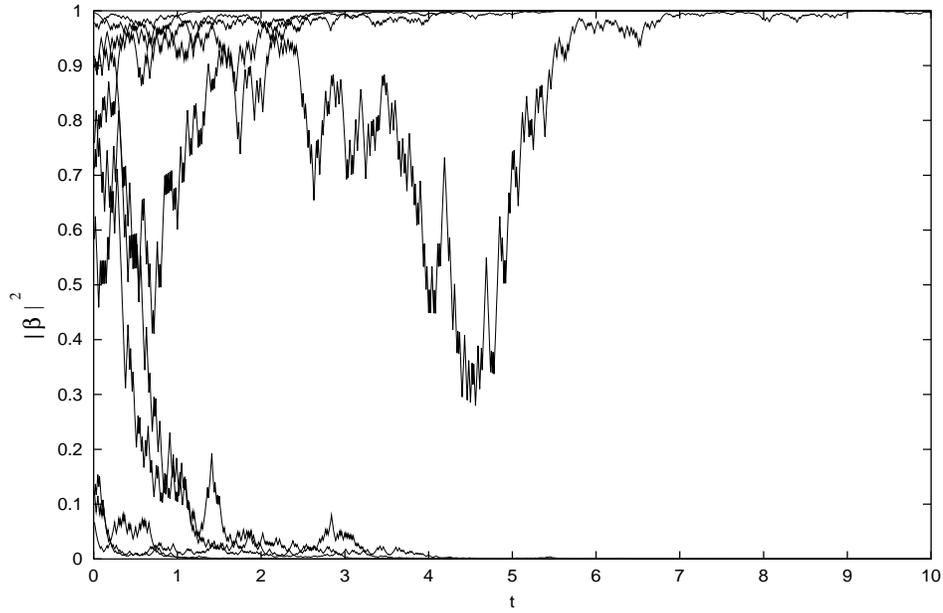, width=5in}
\label{fig4}
\end{center}
\caption{This figure plots $|\beta|^2$ against time for the diffusive
equation (\ref{real_qsd}), with several different initial states and
realizations of the stochastic term.  We see that when the trajectories
approach 0 or 1 they tend to remain there, while in between they diffuse
freely.}
\end{figure}

\end{document}